\documentclass[twocolumn]{aastex631}

\newcommand\aastex{AAS\TeX}

\newcommand{\lya}{$\mathrm{Ly}\alpha\:$}

\received{March 23, 2021}
\revised{July 1, 2021}
\accepted{August 9, 2021}

\submitjournal{ApJ}

\shorttitle{\aastex\ \lya line properties at $z \simeq 3.78$}
\shortauthors{N. Malavasi et al.}

\begin{document}

\title{Lyman Alpha line properties at $z \simeq 3.78$ and their environmental dependence: a case study around a massive proto-cluster}

\correspondingauthor{Nicola Malavasi}
\email{nicola.malavasi@universite-paris-saclay.fr}

\author[0000-0001-9033-7958]{Nicola Malavasi}
\affiliation{Universit{\'e} Paris-Saclay, CNRS, Institut d'astrophysique spatiale, 91405, Orsay, France}
\affiliation{Department of Physics and Astronomy, Purdue University, 525 Northwestern Avenue, West Lafayette, IN 47907, USA}

\author[0000-0003-3004-9596]{Kyoung-Soo Lee}
\affiliation{Department of Physics and Astronomy, Purdue University, 525 Northwestern Avenue, West Lafayette, IN 47907, USA}

\author[0000-0002-4928-4003]{Arjun Dey}
\affiliation{NSF's National Optical-Infrared Astronomy Research Laboratory, 950 N. Cherry Ave., Tucson, AZ 85719, USA}

\author[0000-0001-7689-9305]{Rui Xue}
\affiliation{National Radio Astronomy Observatory, 520 Edgemont Road, Charlottesville, VA 22903, USA}

\author{Yun Huang}
\affiliation{Department of Physics and Astronomy, Purdue University, 525 Northwestern Avenue, West Lafayette, IN 47907, USA}

\author[0000-0003-4747-9656]{Ke Shi}
\affiliation{Department of Astronomy and Institute for Theoretical Physics and Astrophysics, Xiamen University, Xiamen, Fujian 361005, People's Republic of China}

\begin{abstract}
Ly$\alpha$-emitting galaxies (LAEs) are easily detectable in the high-redshift Universe and are potentially efficient tracers of large scale structure at early epochs, as long as their observed properties do not strongly depend on environment. We investigate the luminosity and equivalent width functions of LAEs in the overdense field of a protocluster at redshift $z \simeq 3.78$. Using a large sample of LAEs (many spectroscopically confirmed), we find that the \lya luminosity distribution is well-represented by a \citet{Schechter1976} function with $\log(L^{\ast}/{\rm erg s^{-1}}) = 43.26^{+0.20}_{-0.22}$ and $\log(\phi^{\ast}/{\rm Mpc^{-3}})=-3.40^{+0.03}_{-0.04}$ with $\alpha=-1.5$. Fitting the equivalent width distribution as an exponential, we find a scale factor of $\omega=79^{+15}_{-15}$\AA. We also measured the Ly$\alpha$ luminosity and equivalent width functions using the subset of LAEs lying within the densest cores of the protocluster, finding similar values for $L^*$ and $\omega$. Hence, despite having a mean overdensity more than 2$\times$ that of the general field, the shape of the Ly$\alpha$ luminosity function and equivalent width distributions in the protocluster region are comparable to those measured in the field LAE population by other studies at similar redshift. While the observed Ly$\alpha$ luminosities and equivalent widths show correlations with the UV continuum luminosity in this LAE sample, we find that these are likely due to selection biases and are consistent with no intrinsic correlations within the sample. This protocluster sample supports the strong evolutionary trend observed in the Ly$\alpha$ escape fraction and suggest that lower redshift LAEs can be on average significantly more dusty that their counterparts at higher redshift.
\end{abstract}


\section{Introduction} 
\label{intro}
Studies of galaxy formation, evolution, and cosmology necessarily require that we investigate the high redshift Universe. The class of systems known as Lyman Alpha Emitters \citep[LAEs,][]{PartridgePeebles1967} is particularly helpful to build samples of galaxies in the distant Universe. LAEs are objects whose spectrum presents a fairly strong \lya\ emission line with respect to the Ultra-Violet (UV) continuum. As the \lya emission line is redshifted to optical wavelengths at $z > 2$, it is relatively easy to detect these galaxies in ground-based narrow-band images. Large samples of LAEs have been selected photometrically and confirmed spectroscopically \citep[see, e.g.,][]{CowieHu1998, Kudritzki2000, Ajiki2006, Gronwall2007, Nilsson2009a, Finkelstein2009, Guaita2010, Hayes2010, Ono2010, Ciardullo2012, Ning2020redshift57} at redshift greater than two, with a few samples also in the redshift range $z \sim 6-9$ \citep{Hu2002, Kodaira2003, Iye2006, Willis2006, Ouchi2008, Ono2010, ZhenYaZheng2017, WeidaHu2019, WeidaHu2021}. In recent times several large spectroscopic or photometric surveys have been developed with the explicit aim of specifically targeting LAEs, such as the Hobby-Eberly Telescope Dark Energy Experiment \citep[HETDEX,][]{Hill2008} and its Pilot Survey \citep{Adams2011}, the Large-Area Lyman Alpha survey \citep[LALA,][]{Rhoads2000, MalhotraRhoads2002}, the Multiwavelength Survey by Yale-Chile \citep[MUSYC,][]{Gawiser2006MUSYC}, the Subaru Deep Survey, and Subaru/XMM-Newton Deep Survey \citep[SXDS,][]{Ouchi2003, Ouchi2008}. Large LAE samples resulting from these programs have enabled systematic investigations of their physical properties and their evolutionary connection to the present-day galaxies.

Studies have shown that LAEs are a population of low-mass ($10^{8}-10^{9} M_{\odot}$), star-forming ($1-10\: M_{\odot} \mathrm{yr}^{-1}$) galaxies, abundant in the high-redshift Universe ($z > 2$) \citep{CowieHu1998, Ouchi2003, Ouchi2008, Gawiser2006, Gawiser2006MUSYC, Gawiser2007, Pirzkal2007, Lai2008, Nilsson2007, Nilsson2009a, Finkelstein2009}. They are a numerous class that contribute significantly to the cosmic star-formation rate density (SFRD) at high-redshift \citep[see, e.g.,][]{MadauDickinson2014}.

The complex process that leads to the emission of \lya radiation in galaxies can be studied through the correlations that are present between the strength of the \lya emission line (e.g., its equivalent width, EW) and other properties that trace the various characteristics of a galaxy inter-stellar medium (ISM; e.g., its kinematics, the amount and the distribution of neutral hydrogen gas and dust). For example, \lya luminosity is expected to be correlated with a galaxy's star formation rate (SFR), but this is modulated by the effects of radiative transfer through a complex gaseous and dusty medium. \lya EW is expected to correlate with ISM kinematics, gas column density, and dust reddening \citep{HansenOh2006, Verhamme2006, Verhamme2008, Dijkstra2006, Dijkstra2007, Adams2009, Laursen2010, Zheng2010, Hashimoto2013, CharlotFall1993}, in line with observations \citep[e.g.,][]{Shapley2003}.

The statistics of the LAE population are very helpful in understanding the physical properties of these objects. Since the amount of \lya radiation leaking out of a galaxy strongly depends on the dust distribution and the ISM physics, studying how LAEs in a given volume of space are distributed in \lya luminosity and EW and how these quantities change with redshift and/or environmental density can help understand the evolution and the changes in the physical properties of the ISM of LAEs. The \lya luminosity function (LF) is usually described with a \citet{Schechter1976} function, in which two regimes can be identified: a power-law region at low luminosities; and an exponential decline for the high luminosity tail. The low luminosity power-law region is dominated by faint, mostly low-mass galaxies, and is both sensitive to the physical conditions in the ISM \citep[see, e.g.,][]{Santos2004,Rauch2008} and subject to incompleteness biases. The high luminosity tail is more completely sampled, but is made up of both star-forming galaxies and Active Galactic Nuclei \citep[AGNs, see, e.g.,][]{Sobral2017}. The \lya LF and the EW distribution of LAEs (this latter usually described with an exponential or a Gaussian function) have been explored by many studies \citep[see, e.g.,][]{Gronwall2007, Ouchi2008, Guaita2010, Konno2016, Herenz2019, Fumagalli2021}. Still, the redshift evolution of the parameters describing the two functions is not well constrained, as well as their possible dependence on large scale variations in the environmental density.

In this paper, we study the LAE population in the overdense field of a spectroscopically confirmed proto-cluster at $z \approx 3.78$ and present estimates of its \lya luminosity function and EW distribution. By focusing on a high-density region of space where star-formation in galaxies has likely been accelerated, we can study whether the LAE population properties depend on environment \citep[see, e.g.,][]{Dey2016, Lemaux2018}. Moreover, we study the correlation among the \lya luminosity, EW, and UV continuum luminosity for these sources in order to be able to derive conclusions on the physical properties of their ISM, with particular regard to their dust distribution.

The rest of the paper is organized as follows: in Section \ref{data} we describe the sample used for our work and in Section \ref{lyalfmethod} we describe the procedure we used to measure the \lya luminosity function and EW distribution. In Section \ref{results} we present our results on the LF, while in Section \ref{laenvironment} we describe the environment inhabited by the LAEs in our sample. In Section \ref{ciardulloplots} we analyse the correlation between \lya luminosity, EW, and UV-continuum luminosity. Finally, a discussion of the results and our conclusions are presented in Sections \ref{discussion} and \ref{conclusions}, respectively. Throughout this work we use a concordance cosmology with $\Omega_{m} = 0.27$, $\Omega_{\Lambda} = 0.73$, $\mathrm{H}_{0} = 70\: \mathrm{km} s^{-1} \mathrm{Mpc}^{-1}$.

\section{Data}
\label{data}
In this work, we use a LAE sample at $z \approx 3.78$. This sample has already been thoroughly described in other papers \citep[see][for a detailed description]{Lee2014, Dey2016, Xue2017}. Here we briefly summarize the information which is most important for our analysis.

\begin{deluxetable}{cc}
\tablecaption{Summary of relevant quantities for the LAE sample.\label{obsprop}}
\tablehead{\colhead{Quantity} & \colhead{Value}}
\startdata
Field Center (J2000) & 14:31:35.7 +32:20:05 \\
Redshift & $3.775 - 3.810$ \\
A$_{\rm eff}$ $[\deg^2]$ & 0.526 \\
$N_{\rm LAE}$ & 171 \\
$l_{\rm field}$ [Mpc] & 26.75 \\
$V$ [Mpc$^3$] & $2.255\times 10^5$ \\
\enddata
\tablecomments{A$_{\rm eff}$ is the effective area of the field, $l_{\rm field}$ is the comoving line-of-sight distance corresponding to the narrow-band filter transmission FWHM, and $V$ is the effective comoving volume of the field defined as $A_{\rm eff} \times l_{\rm field}$.}
\end{deluxetable}

Our LAE sample is composed of galaxies in a region encompassing two proto-cluster structures (PC~217.96+32.3, hereafter simply referred to as `PCF') at a redshift of $z \approx 3.786$ \citep[see][]{Lee2014,Dey2016} and those in the general field. This sample covers an area of roughly $1.2 \times 0.6 \deg^{2}$ (i.e., $8\,428 \mathrm{Mpc}^{2}$, see Table \ref{obsprop}) at the southern end of the Bo\"{o}tes field of the NOAO Deep Wide-Field Survey \citep[NDWFS,][]{JannuziDey1999}. The optical data used to define the LAE sample were taken with the Mosaic 1.1 wide-field imaging camera at the Mayall 4m telescope at the Kitt Peak National Observatory (KPNO). 

The three broadband filters ($B_WRI$) are centered at wavelengths of $\lambda_{\rm c} = 4222, 6652, 8118$ \AA$\,$ while the narrowband filter ($WRC4$, KPNO filter number k1024) has a central wavelength of $\lambda_{\rm c} = 5819$ \AA\, and a full-width-at-half-maximum  (FWHM) of $\Delta \lambda$ = 42 \AA. The narrow-band transmission samples the \lya emission at redshifts of $z = 3.775 - 3.810$ corresponding to a line-of-sight distance range of $l_{\rm field} \approx 27$~Mpc. The  volume sampled by these observations is derived as $V = A_{\rm eff} \times l_{\rm field}$ and is reported in Table \ref{obsprop}. 

We isolate LAEs as those galaxies with blue narrow-to-broadband colors ($WRC4-R$), which correspond to an excess of \lya\ emission in the $WRC4$ band. The selection criteria are:
\begin{eqnarray}
\label{pcflaecrit}
(WRC4 - R) < -0.8 \cap {\rm S/N}(WRC4) \ge 7 \cap \nonumber \\
((B_{W} - R) > 1.8 \cup {\rm S/N}(B_{W}) < 2)
\end{eqnarray}
where S/N indicates the signal-to-noise ratio (measured within the isophotal area) in a given band. When these criteria are applied to our sources they produce a sample of 171 objects. Of these, 100 galaxies were observed with the DEep Imaging Multi-Object Spectrograph at the 10m Keck II telescope (DEIMOS, see \citealt{Faber2003}), resulting in the confirmation of 62 galaxies and an estimated spectroscopic success rate of 89\% \citep[as measured by][]{Dey2016}. 

\subsection{Measuring  \lya luminosity and equivalent width}
\label{lyalewmeasurement}
We compute the \lya luminosity and the rest-frame \lya\ equivalent width (EW) of each LAE candidate based on the photometry following the procedure described in \citet{Xue2017}.
Briefly, we define the $Q$ parameter for each passband:
\begin{equation}
Q(z, \beta) = \frac{\int e^{-\tau}(\lambda/\lambda_{0})^{2+\beta} \lambda^{-1} \mathcal{R}(\lambda) d\lambda}{\int  \lambda^{-1} \mathcal{R}(\lambda) d\lambda}
\end{equation}
where $\lambda_0 = 1215.67 (1+z)$~\AA, $\mathcal{R}(\lambda)$ is the  total filter throughput, $\tau$ is the effective optical depth of the IGM as given in \citet{Inoue2014}, and $\beta$ is the slope of the UV continuum ($f_\lambda\propto \lambda^\beta$). Then, the integrated line flux $F_{{\rm Ly}\alpha}$ and continuum flux density $f_{cont}$ are  expressed as:
\begin{eqnarray}\label{ruieqlong}
F_{\mathrm{Ly}\alpha} &=&  \frac{B_{\mathrm{BB}}B_{\mathrm{NB}}(Q_{\mathrm{BB}} f_{\mathrm{AB,NB}} - Q_{\mathrm{NB}} f_{\mathrm{AB,BB}})}{Q_{\mathrm{BB}}B_{\mathrm{BB}} - Q_{\mathrm{NB}}B_{\mathrm{NB}}}   \nonumber \\
f_{\rm cont} &=& \frac{B_{\mathrm{BB}} f_{\mathrm{AB,BB}} - B_{\mathrm{NB}} f_{\mathrm{AB,NB}}}{Q_{\mathrm{BB}}B_{\mathrm{BB}} - Q_{\mathrm{NB}}B_{\mathrm{NB}}}
\end{eqnarray}
The $Q$ parameters and bandpass $B$ for each filter are given in Table 5 of \citet{Xue2017}. In this equation, $f_{\rm AB}$ denotes the flux density in units of erg~s$^{-1}$~cm$^{-2}$~Hz$^{-1}$.  The labels $BB$ and $NB$ refer to the broad-band and the narrow-band which sample the Ly$\alpha$ wavelength, respectively. The monochromatic flux density and the Ly$\alpha$ rest-frame equivalent width are expressed in terms of the above quantities as:
\begin{equation}
f_{\mathrm{AB}} = 10^{-0.4(m_{\mathrm{AB}}+48.6)} = \frac{F_{\mathrm{Ly}\alpha}}{B} + f_{\rm cont}Q
\end{equation}
\begin{equation}
W_0 = \frac{W_{\rm obs}}{1+z} = \frac{\lambda_{0}^{2}}{c} \times \frac{F_{\mathrm{Ly}\alpha}}{(1+z)f_{\rm cont}}
\end{equation}

In computing the line luminosity and EW for each LAE in our sample, we assume that all galaxies lie at the same redshift (taken to be $z=3.790$) for the PCF LAEs. We assume all sources to be at the redshift corresponding to the center of the narrow-band filter. Hence, the dominant source of uncertainty on the \lya luminosity function will be the fact that some sources may be at the filter edge and not at the center of the filter itself (which we take into account by means of our simulations, see Section \ref {simdata}). The filter has a $\mathrm{FWHM} = 42 \mathrm{\AA}$, and therefore the redshift error resulting from our assumption is at most $\pm 0.0173$, which results in a luminosity error of $< 1\%$.

\subsection{Completeness Estimate}
\label{simdata}

Our ability to identify LAEs and measure their line properties depends on the imaging depth, the color selection method we adopt, as well as the redshift of the source. For example, brighter and compact LAEs are easier to be identified than fainter and diffuse ones. For galaxies lying near the edge of the filter transmission curve, the estimated EWs and line luminosities would be underestimated. Here, we describe the extensive image simulations we carry out in order to fully characterize the measurement systematics and their impact on the observed number counts and color distribution. 

We generate a list in which, for each entry, a redshift, UV continuum slope $\beta$, \lya luminosity and EW are randomly chosen from the ranges: $W_0 = 0 - 350$ \AA, $\log{L_{{\rm Ly}\alpha}} = 41.25 - 44.25$, $\beta = -(1.5-2.0)$. As for the redshift range, we assume the range $z = 3.730 - 3.853$. The assumed parameter ranges are much wider than those expected in real data, to ensure that robust statistics can be obtained for the galaxies near the selection cutoffs (in source detection, color selection, and redshift). In particular, the chosen redshift range is much larger than the width of the $WRC4$ filter.

For each entry, we create a spectrum; the base continuum spectrum is generated using the \citet{BC03} stellar population synthesis model assuming a constant star formation history and a population age of 100~Myr. A \citet{Salpeter1955} initial mass function and solar metallicity are assumed. The continuum spectrum is normalized to the input flux density at the rest-frame 1700~\AA; a Ly$\alpha$ emission line normalized to match the line luminosity is added as a Gaussian profile centered at 1215.67~\AA\ and with an intrinsic line width of 3 \AA. Our results are insensitive to the assumed line width as long as they reasonably reproduce the galaxy colors and line widths as observed. As mentioned in \citet{Lee2014}, the line flux is computed independently from the continuum spectrum, to take into account the lower resolution of the \citet{BC03} galaxy templates. The spectrum is then attenuated using the \citet{Inoue2014} prescription to account for IGM absorption by neutral hydrogen. Finally, synthetic photometry is computed by convolving each mock spectrum with the throughputs of the relevant passbands.

Based on the synthetic photometric catalog  we generated above, we carry out image simulations as follows. In each run, we randomly draw 100 entries from the catalog and insert them into the real images. We assume a lognormal distribution for their angular sizes as measured from high-redshift observations \citep{Ferguson2004,Bouwens2004,Law2012,Shibuya2015,Ribeiro2016}. However, our results are insensitive to the morphologies and sizes as all are essentially point sources in our ground-based data. The images are convolved with the appropriate PSF and added to the real data.  A total of $1.5 \times 10^{6}$ sources are simulated in the PCF dataset. Source detection, photometry,  LAE selection, and derivation of line luminosities and EWs are performed in the identical manner as in the real data. 

\begin{figure*}
\begin{minipage}{\linewidth}
\centering
\includegraphics[width = 0.49\linewidth]{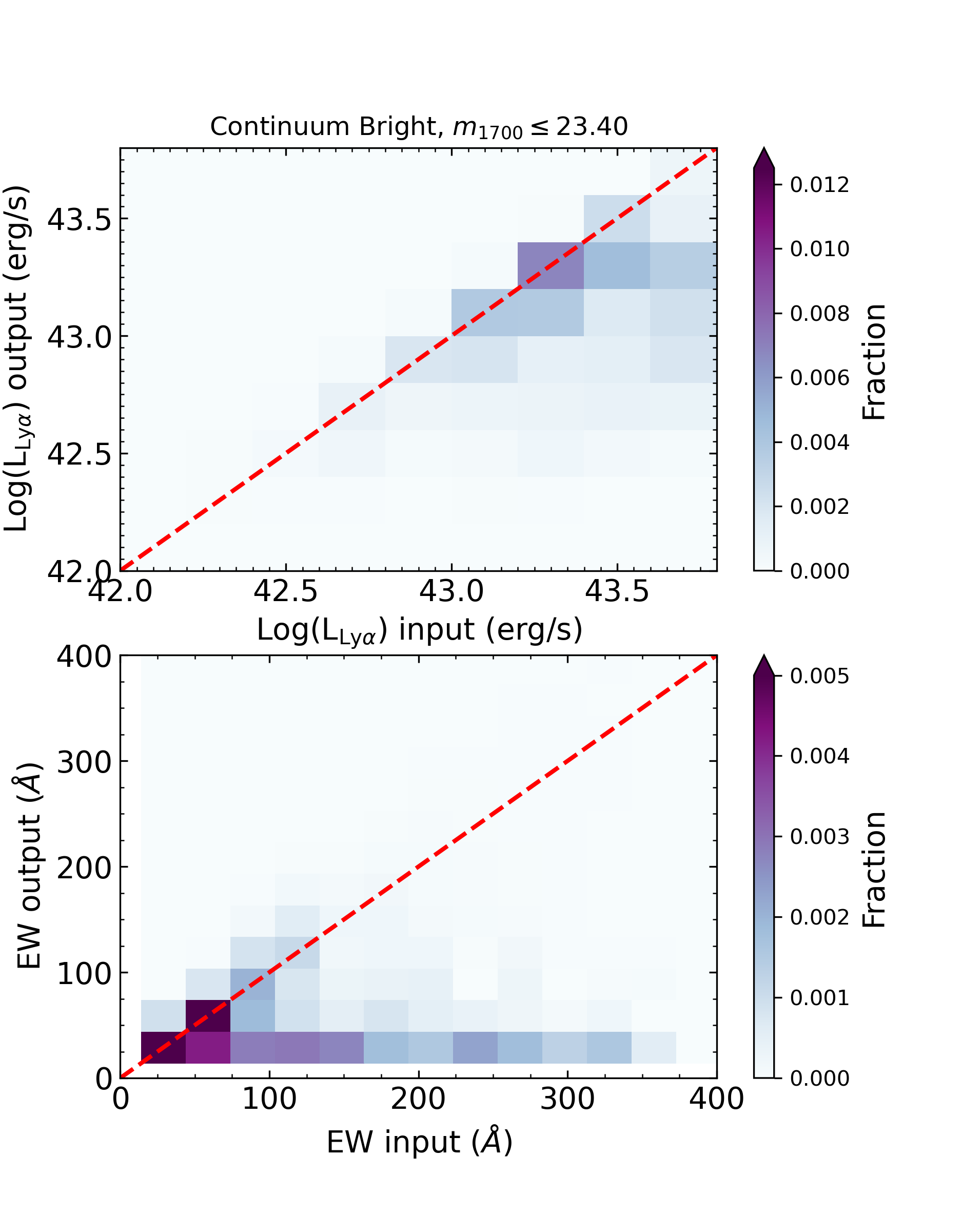}
\includegraphics[width = 0.49\linewidth]{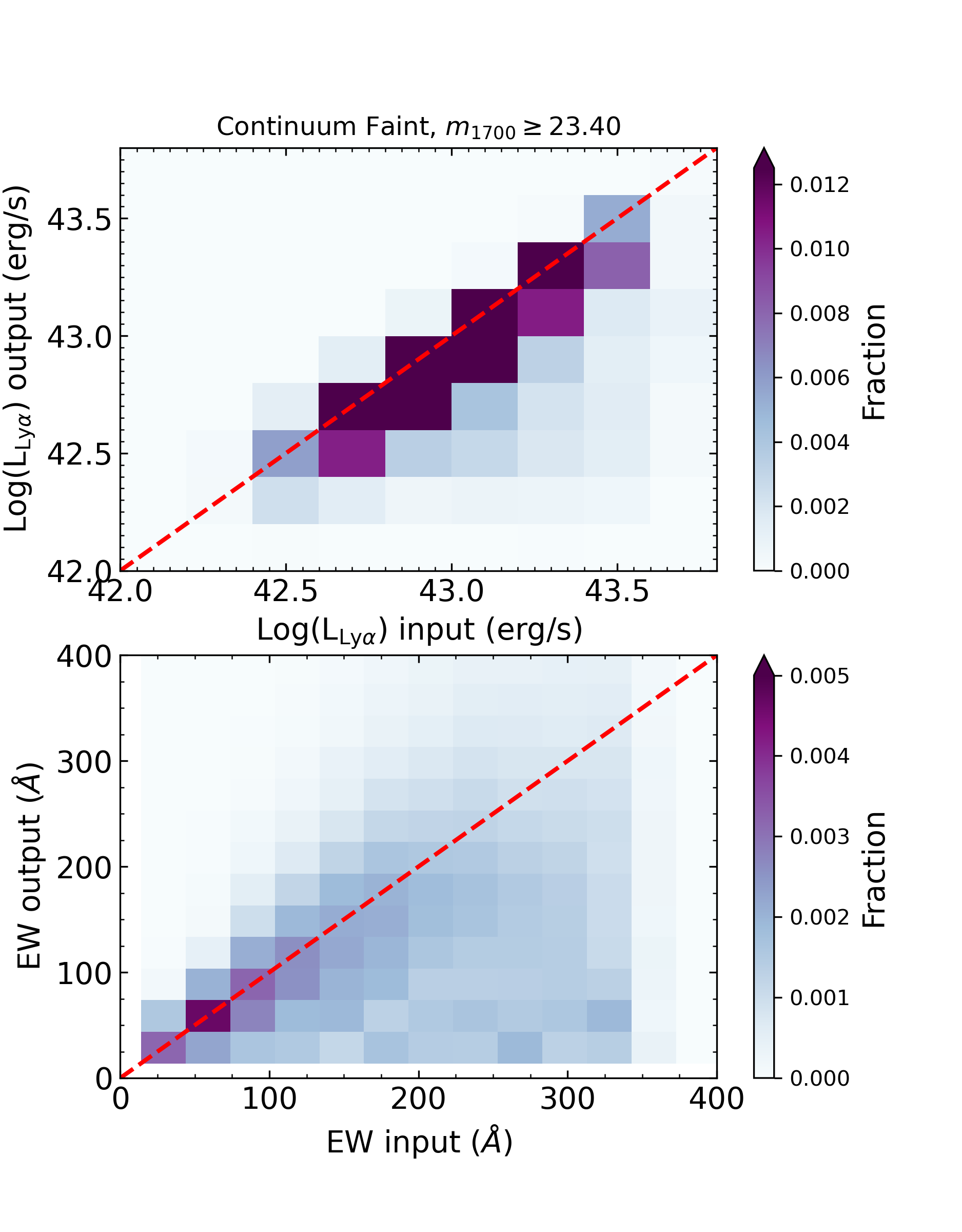}
\end{minipage}
\caption{Relation between input and output \lya luminosity and EW. The color-coding reflects the fraction of all simulated galaxies inside each bin and is consistent between the left and right columns. Left column is for UV-luminous simulated galaxies and right column is for UV-faint ones, where the distinction between the two samples is made considering those below and above the median UV apparent magnitude of the full simulated sample, $m_{1700,\, \mathrm{median}} = 23.40$ AB mag.}
\label{simplots}
\end{figure*}

In Figure \ref{simplots}, we illustrate how both Ly$\alpha$ EWs and luminosities are recovered in our simulations of the PCF dataset when the results are split into the `UV-bright' and `UV-faint' categories ($m_{1700}=23.40$ AB mag is used for the cut). This Figure shows the recovered (output) \lya luminosity and EW as a function of the input ones. It is apparent that UV-bright sources are more likely to be recovered with the correct \lya~luminosity and EW values. In comparison, the same values for UV-faint sources scatter over a wide range.

Using the image simulations we account for the effect of the LAE color selection, detection completeness, and photometric scatter on the observed galaxy statistics. To this end, we define the likelihood that a galaxy with intrinsic luminosity and equivalent width [$L_i$, $W_j$] has of being identified as an LAE and of being observed with the observed quantities [$L_{i'}$, $W_{j'}$] as:
\begin{equation}\label{eqprob}
\mathrm{P}(L_{i'},W_{j'}|L_{i},W_{j}) = \frac{N^{\rm OUT}_{ij}(L_{i'},W_{j'})}{N^{\rm IN}_{ij}}
\end{equation}
where $N^{\rm IN}_{ij}$ is the number of sources simulated in the range [$L_i-\Delta L/2$, $L_i+\Delta L/2$] and [$W_j-\Delta W/2$, $W_j+\Delta W/2$]. The quantity $N^{\rm OUT}_{ij}(L_{i'},W_{j'})$ denotes the number of sources identified as LAEs in the same bin measured with the quantities [$L_{i'}-\Delta L/2$, $L_{i'}+\Delta L/2$] and [$W_{j'}-\Delta W/2$, $W_{j'}+\Delta W/2$]. Throughout this work, all {\it intrinsic} quantities are denoted as unprimed running indices, while all {\it observed} quantities are shown with primed indices.

\begin{figure*}
\begin{minipage}{\linewidth}
\centering
\includegraphics[width = 0.49\linewidth]{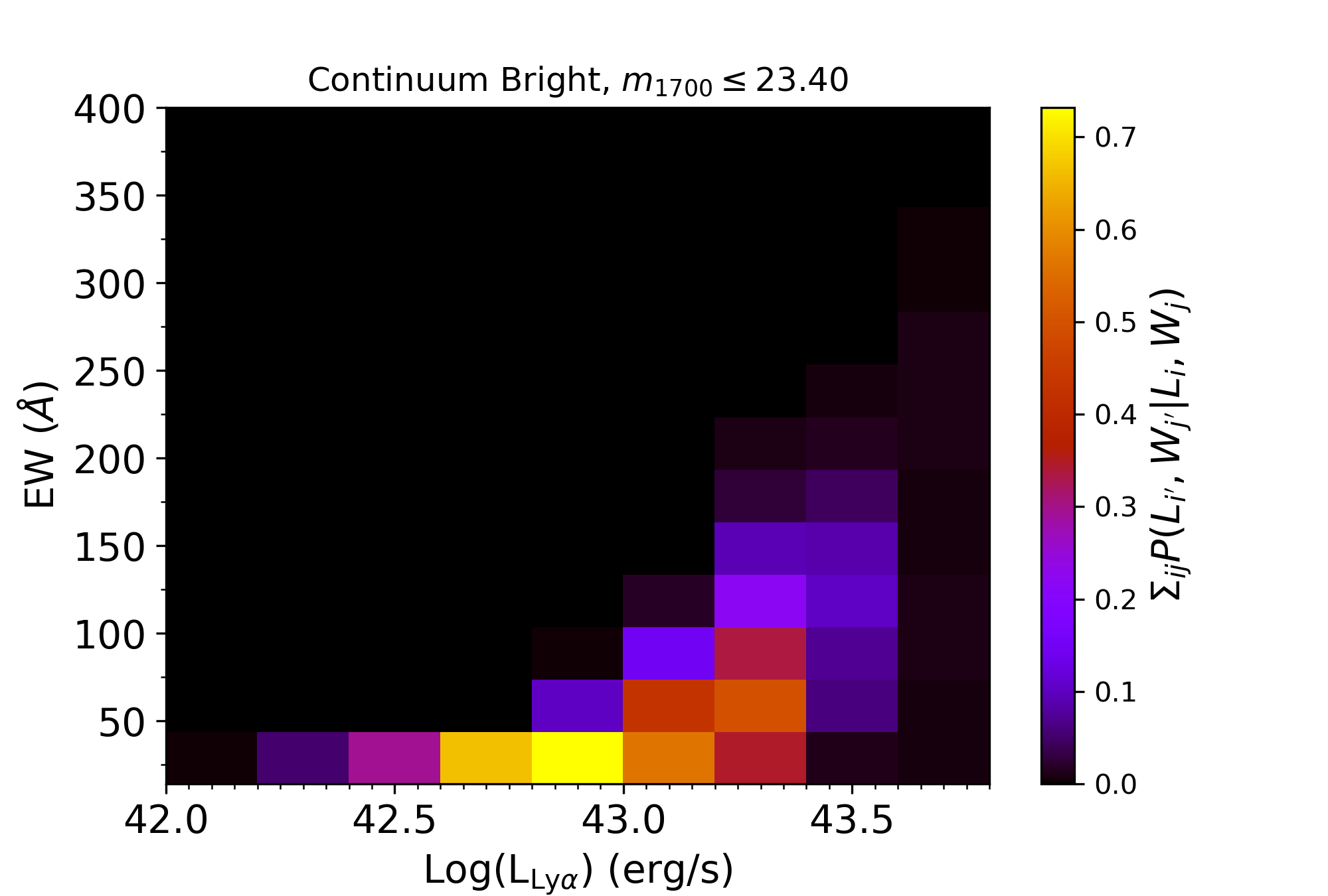}
\includegraphics[width = 0.49\linewidth]{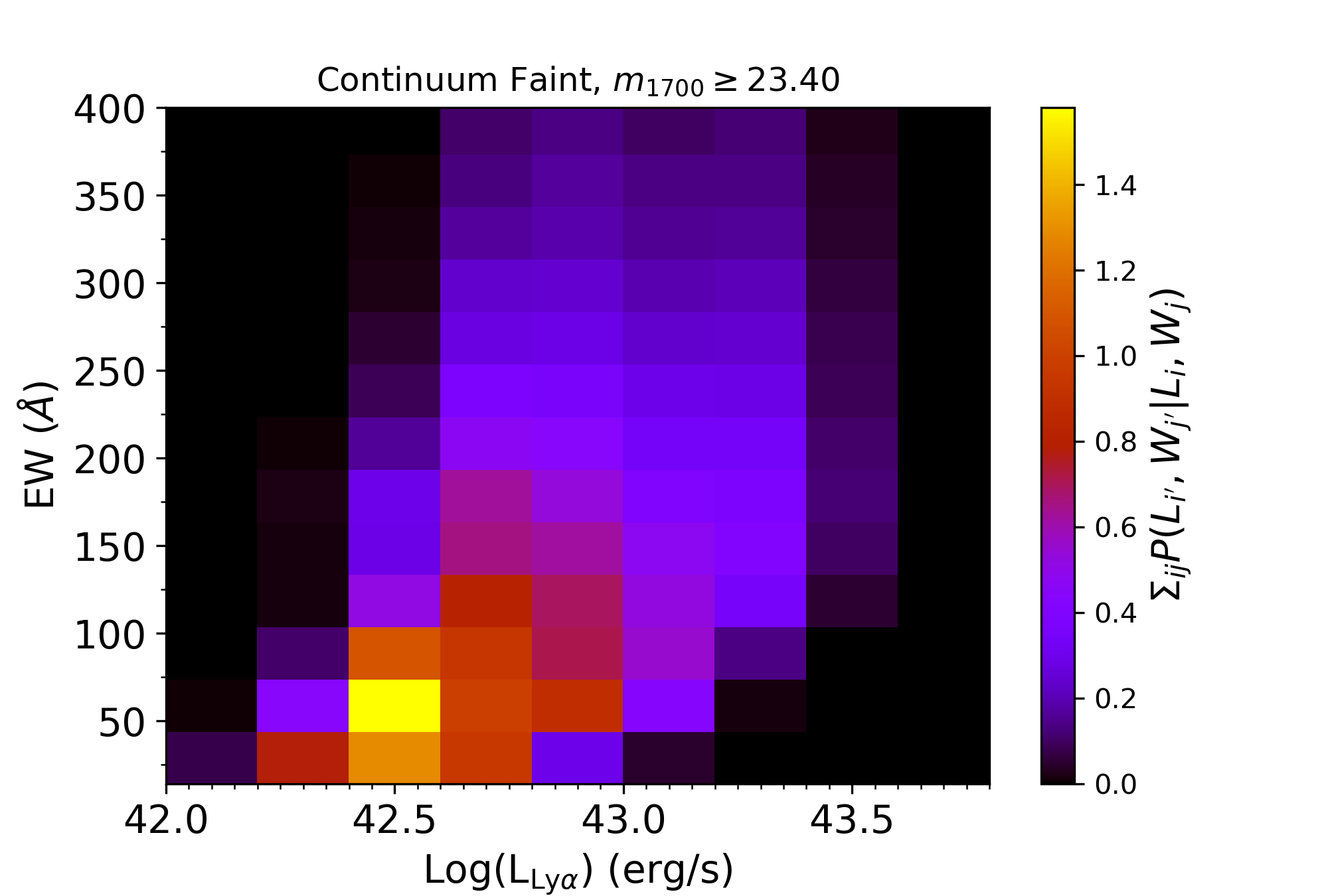}
\end{minipage}
\caption{Two-dimensional distribution of the integrated probability of observing an LAE in a given output \lya$\,$ luminosity-EW bin. The color scale parametrizes the fraction of simulated LAEs that will be observed in a given output \lya$\,$ luminosity-EW bin with respect to the number of LAEs originally in the same input bin. Note that this fraction can be higher than one if more LAEs are scattered in to a bin than out of the bin. Left panel is for UV-luminous simulated galaxies and right panel is for UV-faint ones, where the distinction between the two samples is made considering those below and above the median UV apparent magnitude of the full simulated sample, $m_{1700,\, \mathrm{median}} = 23.40$ AB mag.}
\label{probplots}
\end{figure*}

Figure \ref{probplots} reports an estimate of the ratio of observed to expected LAE counts in the \lya-EW space. The distributions shown in Figure \ref{probplots} are obtained when Equation \ref{eqprob} is summed over all intrinsic \lya luminosity and EW values ($i,j$ indices) and shown as a function of the observed \lya luminosity and EW values ($i',j'$ indices) for continuum bright and continuum faint sources. The sum of the probability from Equation \ref{eqprob} over the intrinsic values describes how the detected fraction at a given \lya luminosity depends on the photometric uncertainties. For the majority of continuum bright sources at least 30\% or more of the sources in a given intrinsic \lya~luminosity and EW bin are observed as belonging to the same bin. Moreover, the distribution in the observed space of the bins for which the probability of observing an LAE with the correct \lya luminosity and EW is high is rather compact, restricted to a smaller range of observed \lya luminosities and EWs than in the continuum faint case.

For continuum faint sources, several bins are also present where we observe more galaxies than expected based on the intrinsic number (i.e., an observed fraction greater than one). This is due to galaxies being scattered in  to a given \lya luminosity and EW bin from neighbouring ones due to photometric uncertainties. Moreover for continuum faint sources, the number of bins with a very low fraction of observed simulated galaxies is larger and scattered also to higher EW values and \lya luminosities.

\section{Methodology}
\label{lyalfmethod}

In this work, we aim to determine the intrinsic distributions of \lya luminosity and equivalent width. Following the convention in the literature \citep[e.g.,][]{Gronwall2007, Guaita2010, Blanc2011, Ciardullo2012}, we model the luminosity function and EW function as a \citet{Schechter1976} function and an exponentially declining function, respectively. They are expressed as:
\begin{equation}\label{schechter}
\Phi(L_{i}) = \ln(10) \log(\phi^{\ast}) e^{-\frac{L_{i}}{L^{\ast}}}  \left(\frac{L_{i}}{L^{\ast}} \right)^{\alpha+1}
\end{equation}
\begin{equation}\label{gaussian}
\Psi(W_{j}) =  \psi^{\ast} e^{-W_{j}/{\omega}}
\end{equation}
In the above equations, both functions represent comoving number densities in a unit log luminosity $[\log(L_{i}), \log(L_{i})+\Delta \log(L)]$ and EW bin $[W_j,W_j+\Delta W]$. $L^*$ is the characteristic luminosity while $\alpha$ and $\phi^*$ denote the faint-end slope and the normalization parameter, respectively. The equivalent width function (EWF) is a single-parameter function characterized by the width $\omega$ as the normalization itself $\psi^\ast$ is tied to the LF.

Given the true LF and EWF, we can predict the expected luminosity and EW distributions `as observed' by essentially convolving the observational selection and biases with these quantities as:

\begin{eqnarray}
N_{\rm exp}(L_{i'},W_{j'}) = \nonumber \\
\sum_{i,j} \mathrm{P}(L_{i'},W_{j'}|L_{i},W_{j})\: \Phi(L_{i})\: \Delta \log(L)\:  \Psi(W_{j})\:  V
\end{eqnarray}
where $V$ is the comoving volume of the survey as listed in Table \ref{obsprop}. By comparing this expectation with the observed distribution, one can determine the best-fit parameters  $L^*$, $\alpha$, $\phi^*$, and $\omega$. 

To this end, we employ a maximum likelihood method similar to that presented in \citet[][]{Bouwens2007}. The log-likelihood is defined as:
\begin{equation}
p_{i'j'} \equiv N_{\rm obs}(L_{i'},W_{j'}) \ln \left( \frac{N_{\rm exp}(L_{i'},W_{j'})}{\sum_{m,n} N_{\rm exp}(L_{m},W_{n})} \right)
\end{equation}
where the sum at the denominator is performed for all $m, n$ while the $p_{i'j'}$ values are computed only on the set of indices $\mathcal{A} = \{i', j' : N_{\rm obs}(L_{i'},W_{j'}) \ne 0\}$. The log-likelihood is then computed as $\mathcal{L} = \sum_{i', j' \in \mathcal{A}} p_{i'j'}$. The best-fit solution is the one that maximizes the log-likelihood $\mathcal{L}$.

Our fitting method departs from the convention in the literature in that we make full use of the 2D distribution rather than fitting the \lya\ luminosity and EW functions, separately. If the two are related in a known fashion, the methodology can be easily generalized to fit such a model. In the case of no intrinsic relation between the two quantities, our method should yield a result equivalent to the conventional method. While we proceed assuming that there is no correlation, in Section~\ref{ciardulloplots}, we return to the possibility that the two quantities are linked. 

A wide range of parameters are considered in the fitting procedure, ranging in $\log(L^{\ast}/(erg \cdot s^{-1})) = 40 - 45$, $\alpha=-(1.50-2.25)$, and $\omega=(1-700)$~\AA, respectively. Using the best-fit parameters, we compute $N_{\rm best}(L_{i'},W_{j'})$ and define the expected number densities per log-luminosity and per EW as:
\begin{eqnarray}\label{lfeqbestfit}
\Phi_{\rm exp}(L_{i'}) &=& \frac{\sum_{j'} N_{\mathrm{best}}(L_{i'},W_{j'})}{\Delta \log(L) \cdot V}~; \nonumber \\
\Psi_{\rm exp}(W_{j'}) &=& \frac{\sum_{i'} N_{\mathrm{best}}(L_{i'},W_{j'})}{\Delta W \cdot V}
\end{eqnarray}
and similarly for the {\it observed} number densities, substituting $N_{\rm best}(L_{i'},W_{j'})$ with $N_{\rm obs}(L_{i'},W_{j'})$.

Finally, the normalization parameters for both functions ($\phi^{\ast}$, $\psi^{\ast}$) can be obtained as:
\begin{equation}
\phi^{\ast} = \frac{\sum_{i'} \Phi_{\rm obs}(L_{i'})}{\sum_{i'} \Phi_{\rm exp}(L_{i'})}; ~~~~\psi^{\ast} = \frac{\sum_{j'} \Psi_{\rm obs}(W_{j'})}{\sum_{j'} \Psi_{\rm exp}(W_{j'})}
\end{equation}
The confidence levels of the parameters are estimated assuming that the likelihood ratio $\Lambda = -2 (\mathcal{L}-\mathcal{L}_{\rm max})$ obeys a $\chi^{2}$ distribution with a number of degrees of freedom equal to the number of free parameters in the fit.

\section{Ly$\alpha$ Luminosity and Equivalent Width Functions}
\label{results}

\begin{figure}
\includegraphics[width=\linewidth, trim=0cm 1cm 0cm 1cm, clip=true]{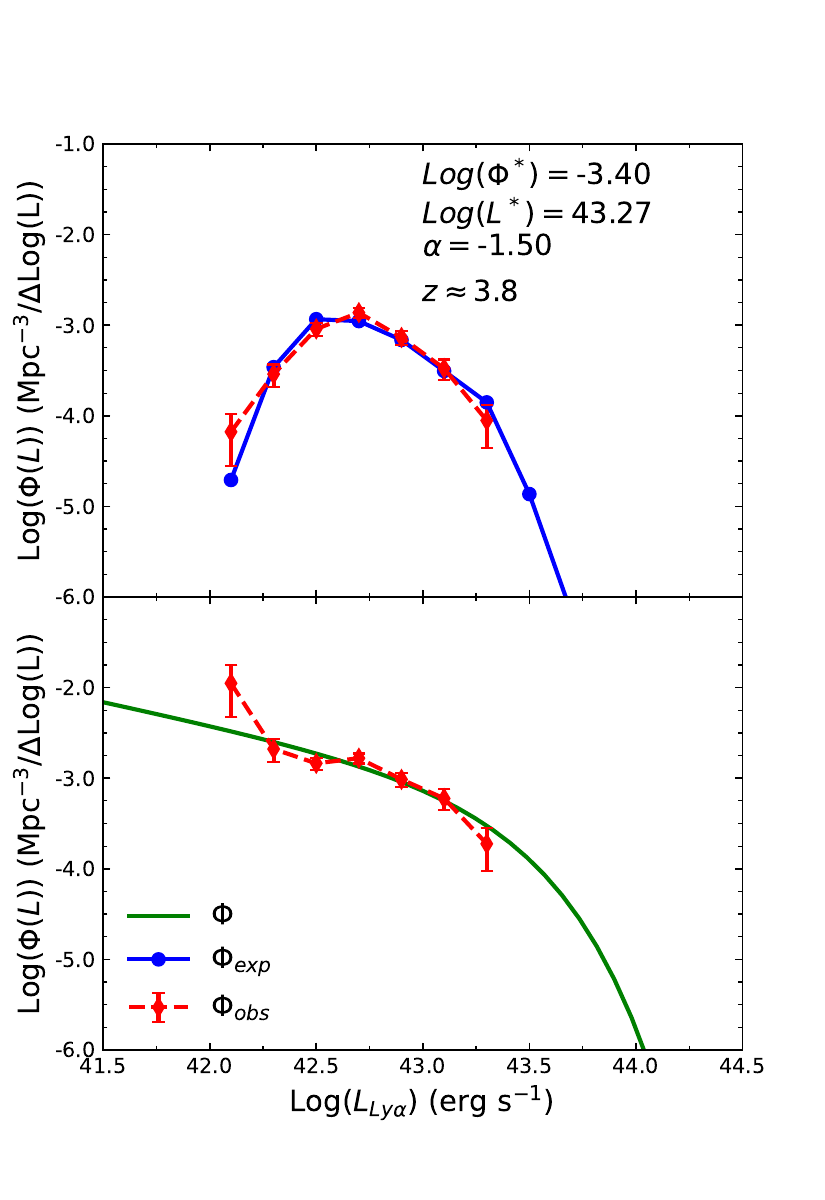}\\
\caption{The comoving number density of LAEs are shown as observed (top) and as corrected for measurement and selection biases (bottom). {\it Top:} both data (red) and best-fit model (blue) include measurement and selection biases as described and characterized in Section~\ref{simdata}. {\it Bottom:} observed densities corrected for incompleteness are shown in red, the green line represents the best-fit Schechter function.}
\label{lf}
\end{figure}

Following the procedure described in Section~\ref{lyalfmethod}, we measure the LF and EWF using the full LAE sample at $z \approx 3.8$. In Figure~\ref{lf}, we show our measurements together with the best-fit models. In the top panel, both data and models show the comoving number densities `as observed' while, in the bottom panel, the same quantities are shown as `intrinsic'. The two are related through Equation~10.

We are unable to place robust constraints on the faint-end slope, $\alpha$. Formally, the best-fit $\alpha$ values is $-1.5$. This value is in a reasonable agreement with other measures within uncertainties \citep[e.g.,][]{Sobral2017}. However,  all $\alpha$ values listed in Table~\ref{bestfitparams} provide similarly good fits to the data as illustrated by the $\mathcal{L}_{\rm max}$ values. We instead opt to fix $\alpha$ to five different values, linearly spaced in the range $\alpha=-(1.50-2.25)$, and determine $L^*$, $\Phi^*$ and $\omega$ for each case. The results are tabulated in Table~\ref{bestfitparams} listed under `full sample'. For the ease of  comparison with other results in the literature, we fix the faint-end slope to $\alpha =-1.50$ for the $z \approx 3.8$ sample. 

\begin{deluxetable*}{cccccc}
\tablecaption{Best-fit parameters for the rest-frame \lya luminosity and equivalent width functions\label{bestfitparams}.} 
\tablehead{
\colhead{Field} & \colhead{$\alpha$} & \colhead{$\log{L^{\ast}}$~[erg~s$^{-1}$]} & \colhead{$\log{\phi^{\ast}}$~[Mpc$^{-3}$]} & \colhead{$\omega$~[\AA]} & \colhead{$\mathcal{L}_{max}$}}
\startdata
{	} & $-1.50$ & $43.26_{-0.22}^{+0.20}$ & $-3.40^{+0.03}_{-0.04}$ & $79_{-15}^{+15}$ & $-620.4$ \\
{	} & $-1.65$ & $43.37_{-0.23}^{+0.34}$ & $-3.57^{+0.03}_{-0.04}$ & $79_{-14}^{+16}$ & $-620.9$ \\
{PCF} & $-1.70$ & $43.37_{-0.20}^{+0.47}$ & $-3.60^{+0.03}_{-0.04}$ & $79_{-14}^{+16}$ & $-621.1$ \\
{171 LAEs	} & $-1.75$ & $43.47_{-0.26}^{+0.53}$ & $-3.74^{+0.03}_{-0.04}$ & $79_{-14}^{+17}$ & $-621.2$ \\
{	} & $-2.00$ & $44.08_{-0.62}^{+0.92}$ & $-4.61^{+0.03}_{-0.04}$ & $79_{-13}^{+19}$ & $-622.3$ \\
{	} & $-2.25$ & \ldots\tablenotemark{a} & $-6.11^{+0.03}_{-0.03}$ & $79_{-11}^{+22}$ & $-624.3$ \\
\hline
{	}        & $-1.50$ & $43.26_{-0.31}^{+0.97}$ & $-$ & $72_{-21}^{+40}$ & $-149.2$ \\
{	}        & $-1.65$ & $43.47_{-0.43}^{+1.53}$ & $-$ & $72_{-20}^{+42}$ & $-149.2$ \\
{PCF, HD}    & $-1.70$ & $43.47_{-0.41}^{+1.53}$ & $-$ & $72_{-20}^{+43}$ & $-149.2$ \\
{43 LAEs}        & $-1.75$ & $43.57_{-0.47}^{+1.43}$ & $-$ & $72_{-20}^{+43}$ & $-149.2$ \\
{	}        & $-2.00$ & \ldots\tablenotemark{a} & $-$ & $72_{-19}^{+47}$ & $-149.3$ \\
{	}        & $-2.25$ & \ldots\tablenotemark{a} & $-$ & $79_{-25}^{+43}$ & $-149.8$ \\
\hline
{	}            & $-1.50$ & $43.26_{-0.24}^{+0.59}$ & $-$ & $72_{-17}^{+24}$ & $-258.8$ \\
{	}            & $-1.65$ & $43.47_{-0.35}^{+1.53}$ & $-$ & $72_{-16}^{+25}$ & $-259.1$ \\
{PCF, Structure} & $-1.70$ & $43.57_{-0.43}^{+1.43}$ & $-$ & $72_{-16}^{+26}$ & $-259.2$ \\
{72 LAEs}            & $-1.75$ & $43.57_{-0.39}^{+1.43}$ & $-$ & $72_{-16}^{+26}$ & $-259.3$ \\
{	}            & $-2.00$ & \ldots\tablenotemark{a} & $-$ & $72_{-16}^{+29}$ & $-260.0$ \\
{	}            & $-2.25$ & \ldots\tablenotemark{a} & $-$ & $72_{-15}^{+31}$ & $-261.3$ \\
\enddata
\tablenotetext{a}{In this case the fitting procedure is not able to constrain the $L^{\ast}$ parameter.}
\tablecomments{Columns are: the sample for which the LF and EWF have been computed (Field, with PCF referring to the full LAE sample, PCF HD referring to the LAEs in high densities defined through the Delaunay tessellation, and PCF Structure referring to LAEs within spectroscopically identified structures), the $\alpha$, $L^{\ast}$, and $\phi^{\ast}$ parameters of the LF, the $\omega$ parameter of the EWF, and the maximum likelihood parameter of the fit ($\mathcal{L}_{max}$). Below the name of the sample is the number of LAEs used to constrain the fit.}
\end{deluxetable*}

\begin{figure}
\includegraphics[width=\linewidth]{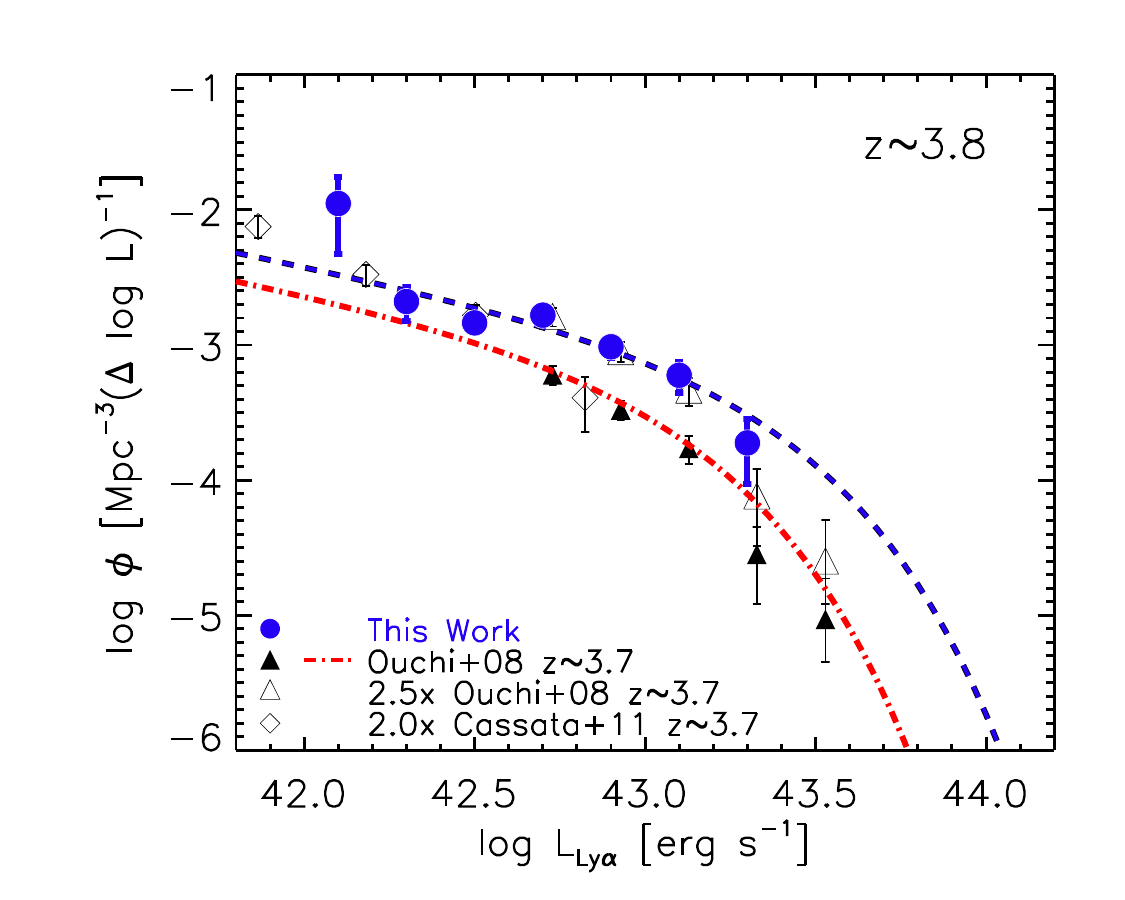}
\caption{Comparison of measured Ly$\alpha$ LFs at $z \approx 3.8$ with those reported by \citet{Ouchi2008} and \citet{Cassata2011}. Dashed lines represent the best-fit Schechter function as tabulated in Table \ref{bestfitparams}. Data from the literature have been corrected to the cosmology adopted in this paper. We find that the overall normalization of our LAE sample is a factor of $\sim 2.0-2.5$ higher than that in the average field at the bright end.}
\label{lf_compare}
\end{figure}

In Figure~\ref{lf_compare}, we show our measurements together with those at $z \approx 3.7$ \citep{Ouchi2008, Cassata2011}. The bright end of our LF is not well sampled, due to a small sample size, leading to large uncertainties in characteristic luminosity $L^*$. Nevertheless, there is a clear enhancement of the overall amplitude of the LF (of $\sim 0.3$ dex) compared to the \citet{Ouchi2008} determination based on field LAEs at the same redshift, in excellent agreement with our earlier estimate that the observed number of LAEs in the entire survey field is nearly twice the average expected in the field environments \citep{Lee2014}. Although our high-luminosity end is not well sampled, there is also an indication for the shape to be different from what measured by \citet{Ouchi2008}, with an enhancement in our LF with respect to the one of field LAEs. If confirmed, this could point towards an enhancement and an acceleration of galaxy formation in proto-cluster cores, as suggested, e.g., by \citet{Shimakawa2018, Ito2020}, albeit through analyses performed with different kinds of data sets.

\begin{figure}
\includegraphics[width=\linewidth]{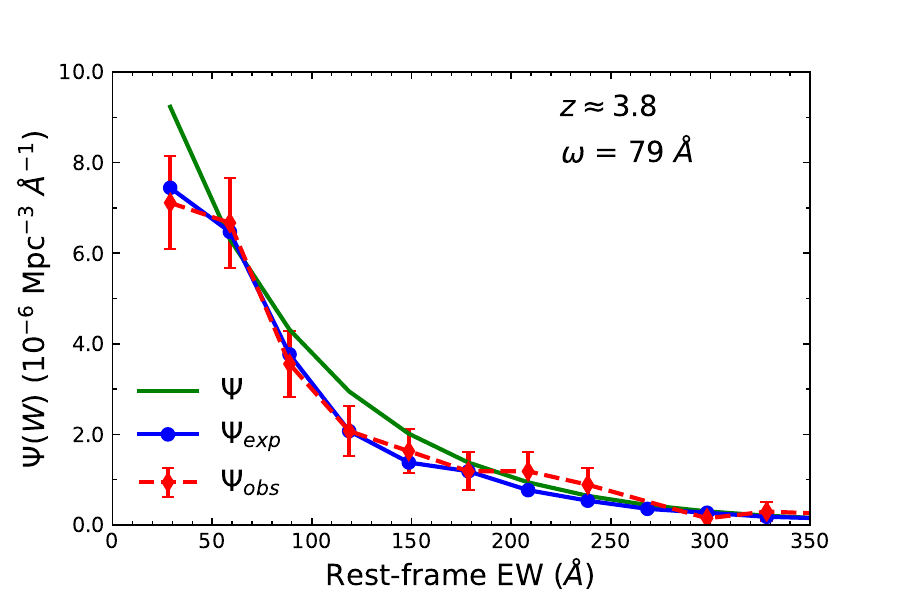}
\caption{Distribution of Ly$\alpha$ equivalent widths are shown for the data in red for our LAE sample. The best-fit EWF is shown in green, together with the expected EW distribution in blue given the selection effects and photometric scatter (see Equation~\ref{eqprob}).}
\label{ewf}
\end{figure}

In Figure~\ref{ewf}, we show the observed EW distribution for our sample along with the best-fit intrinsic EWF (green). The EW distribution is well fitted by an exponential function which shows how the majority of sources have $\mathrm{EW} < 250$ \AA$\:$ with a tail extending up to 350 \AA$\:$ and more.

\begin{figure*}
\centering
\includegraphics[width=5in]{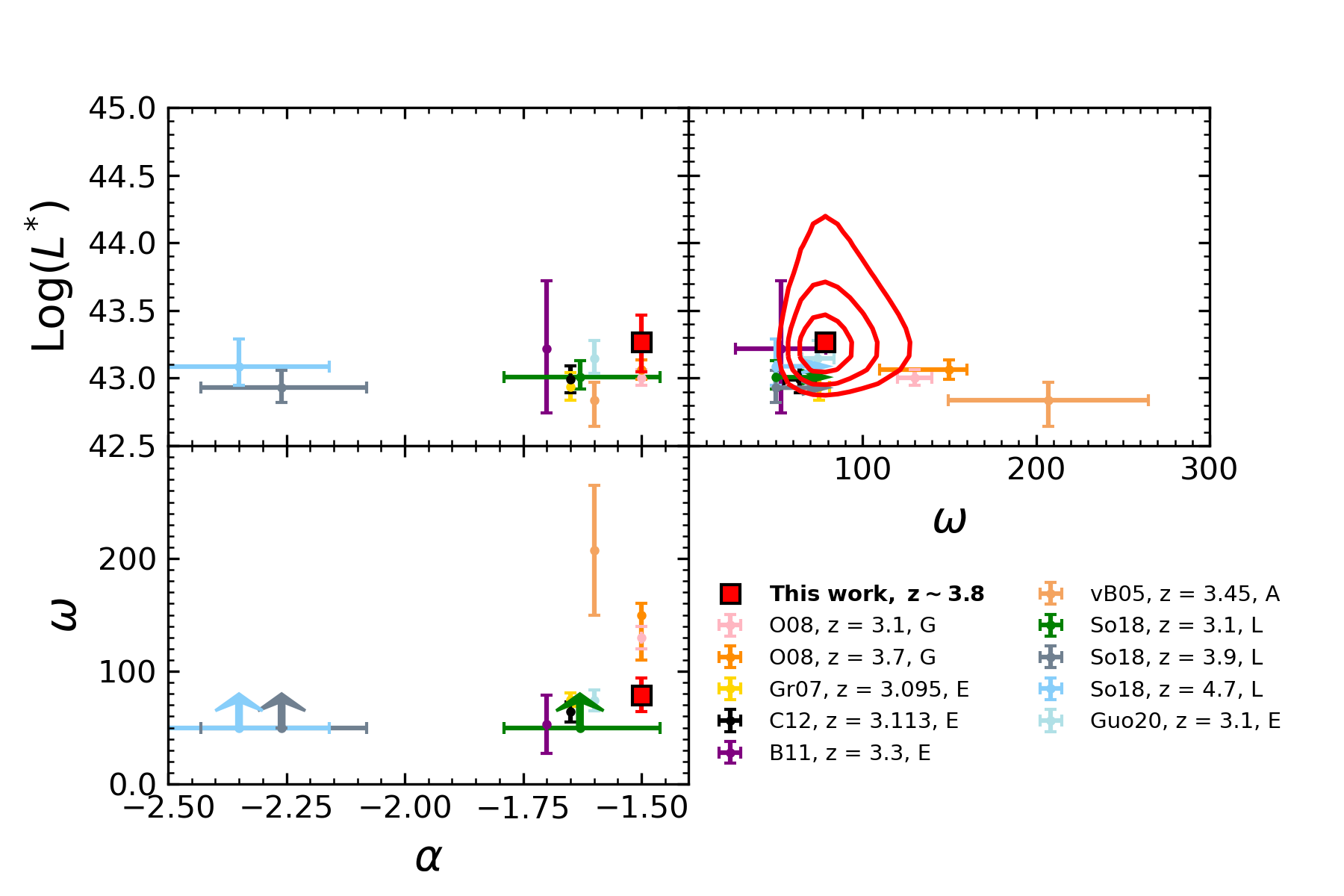}\\
\caption{
Best-fit LF and EWF parameters for our LAE sample are shown in red. The  contours represent $1\sigma$, $2\sigma$, and $3\sigma$ confidence levels. The faint-end slope $\alpha$ is fixed to  the values assumed by \citet{Ouchi2008}, which is $-1.50$ for our $z \approx 3.8$ sample. Other colors represent the measurements from the existing literature at $z=3-5$: \citet[][O08]{Ouchi2008}, \citet[][Gr07]{Gronwall2007}, \citet[][C12]{Ciardullo2012}, \citet[][B11]{Blanc2011}, \citet[][vB05]{vanBreukelen2005}, \citet[][So18]{Sobral2018}, \citet[][Guo20]{Guo2020}. Letters in the legend refer to they way in which $\omega$ has been computed, whether it is from a Gaussian or Exponential fit to the distribution (G or E, respectively), the average of the distribution (A), or a lower limit (L).} The limits are displayed as arrows whenever appropriate. These literature data are corrected for different cosmology whenever applicable.
\label{contours}
\end{figure*}

As illustrated in Figure~\ref{contours}, our measurements are in good agreement with those in the literature at similar redshift \citep{vanBreukelen2005, Gronwall2007, Ouchi2008, Blanc2011, Ciardullo2012, Sobral2018, Guo2020} and are consistent with the limits set by \citet{Sobral2018}. In the case of the EWF parameter $\omega$, \citet{Ouchi2008} parametrised the EW distribution with a Gaussian function, whereas we model it with an exponential function. The value of $\omega$ we report in Figure~\ref{contours} for \citet{Ouchi2008} is therefore the standard deviation of a Gaussian (and not the e-folding of an exponential), therefore it is in principle not directly comparable to our measurement of $\omega$. However, we checked that values extracted from a range of Gaussian functions centered on zero and with standard deviations corresponding to the range covered by our parameter space are also well fitted by exponential functions with e-folding parameters ranging over the same range. Therefore, our measurement for $\omega$ and the one from \citet{Ouchi2008} can be directly compared. The contours in the $L^*$-$\omega$ space represent the $1\sigma$, $2\sigma$, and $3\sigma$ confidence levels; it is evident that the small size of the PCF $z \approx 3.8$ sample translates to large uncertainties in particular for the determination of the characteristic luminosity $L^*$.

\section{The impact of the environment on \lya\ emitters}
\label{laenvironment}

As outlined in Section~\ref{data}, our sample contains 54 spectroscopically confirmed members of two $z \approx 3.78$ structures \citep[PC217.96+32.3-C and PC217.96+32.3-NE, hereafter: PCF, see][]{Lee2014, Dey2016}. The estimated total masses of these structures are $\approx 1 \times 10^{15}M_\odot$ and $6 \times 10^{14}$ $M_{\odot}$, respectively. 

We quantify the environment traced by the LAEs at $z \approx 3.78$ by performing a two-dimensional Delaunay tessellation \citep{Marinoni2002, vandeWeygaert1994, Zaninetti1990},
a geometrical dual operation of the Voronoi Tessellation \citep[see][]{Marinoni2002}. The line-of-sight distance sampled by the narrow-band filter $WRC4$ is $\approx 20$~Mpc, roughly comparable to that of the angular extent of the LAE overdensity. We use a 2D approach as the precise positions of individual LAEs are unknown in the line-of-sight direction, due to the fact that \lya line centroids are typically redshifted with respect to systemic redshift, with the offset sensitive to the geometry, motion and physical properties of interstellar gas.

The 2D plane is divided in a series of triangles having  LAEs as vertices  with a property that, for every triangle, the enclosing circle uniquely defined by its vertices does not contain any other galaxy. Unlike in the case of Voronoi tessellation, the polygons that tessellate the space are all triangles. Each galaxy is connected with a number of triangles $N_{T}$ each of area $A_{t}$ with $t = 0, 1, \ldots, N_{T}$. The surface density of each LAE is determined as an average of the inverse of the areas of the triangles:

\begin{equation}
\Sigma_{\rm D} = \frac{1}{N_{T}} \sum_{t}\frac{1}{A_{t}}
\end{equation}
The surface overdensity is then defined as:
\begin{equation}
\delta = \frac{\Sigma_{\rm D}-\langle \Sigma \rangle}{\langle \Sigma \rangle}
\end{equation}
where $\langle \Sigma \rangle$ is the mean LAE density computed by dividing the total number of LAEs by the effective field area, which is listed in Table~1. 

The use of the Delaunay tessellation is a slightly different approach than the one adopted by \citet{Dey2016}, which relied on Voronoi tessellation. In this case, we wanted to take advantage of the fact that the measurement of the density field through the Delaunay tessellation allows for a slight smoothing of the density at the position of each galaxy, by means of the average of the areas of the triangles which connect it to the neighbouring galaxies. In this way, the density at the position of each galaxy is measured taking into account the density at the position of the neighbouring ones, as the same triangle will contribute to the density measurement of multiple LAEs. 

\begin{figure}
\includegraphics[width = \columnwidth, trim = 0cm 4cm 0cm 1cm, clip = true]{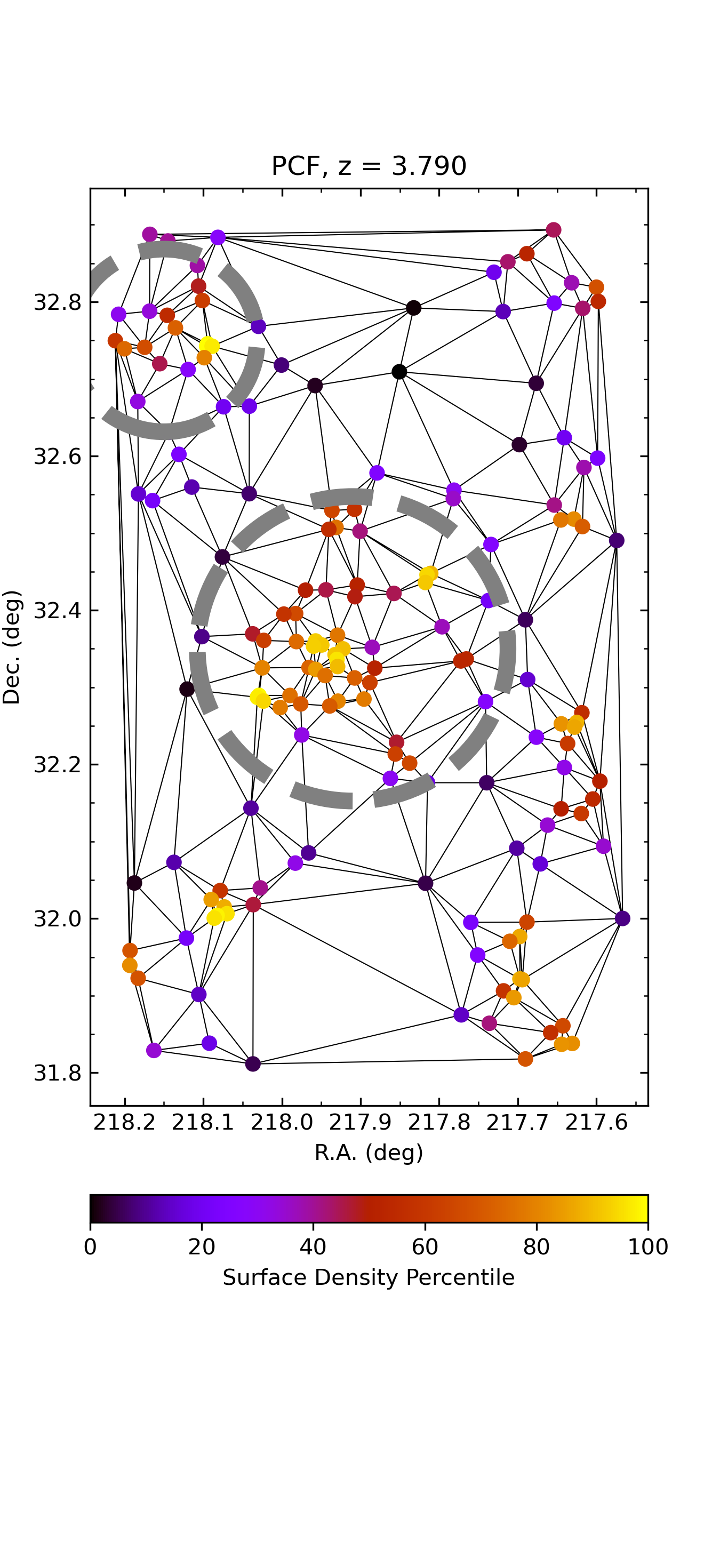}
\caption{Delaunay tessellation of the $z \approx 3.8$ sample. Black lines show the tessellation polygons while circles represent the positions of candidate LAEs in the field. Galaxies are colour-coded according to the percentile of the estimated LAE surface density. Two grey circles mark the positions of PC217.96+32.3C and PC217.96+32.3-NE \citep{Lee2014}, and  have radii 25 Mpc and 15~Mpc (comoving), respectively.}
\label{tespcf}
\end{figure}

In Figure~\ref{tespcf}, we show the Delaunay tessellation of the PCF field, where LAEs are color-coded according to the percentile of the surface density. The positions of the two spectroscopically confirmed proto-clusters are marked by dashed grey circles (with radii of 25 Mpc and 15~Mpc, comoving, for the central and NE structure, respectively); the majority of the LAEs therein are at the 80$^{\rm th}$ percentile in surface overdensity. The radii of the overdensities are very approximate and are based on visual estimates \citep[see][]{Lee2014}. We define the high-density LAE sample in two different ways: one according to the tessellation map, and the other to the spectroscopic overdensity. These samples will be referred to as ``HD'' and ``Structure'' sample, respectively. The ``HD'' sample is defined as those LAEs with a density greater than the 75th percentile of the density distribution (43 LAEs belong to this sample), the ``Structure'' sample as those LAEs within a radius of 25 and 15 comoving Mpc from the positions of the central and NE overdensities, respectively (72 LAEs belong to this sample).

We determine the best-fit LF and EWF parameters for the high-density LAE samples. The faint-end slope is fixed to the same values as the full sample. The results are listed in Table~\ref{bestfitparams} under ``Structure'' for the spectroscopy-defined overdensity sample and ``HD'' for the tessellation-based sample.  At $z=3.78$, the characteristic luminosity and EWF width are (assuming $\alpha=-1.5$): $\log{L^\ast}=43.26^{+0.20}_{-0.22}$, $\omega =79^{+15}_{-15}~$\AA\ for the full sample, $\log{L^\ast}=43.26^{+0.59}_{-0.24}$, $\omega=72^{+24}_{-17}$~\AA~ for the ``HD'' sample, and $\log{L^\ast}=43.26^{+0.97}_{-0.31}$, $\omega=72^{+40}_{-21}$~\AA~ for the ``Structure'' sample, respectively. All parameters obtained for the high-density LAEs are consistent with those of the full sample within errors. It appears that there is no significant difference in the formation process of LAEs in the average and overdensity field.

We chose to compare the LF and EWF in overdensities with the total LF and EWF because we wanted to check whether the LAE population in very dense environments showed any difference in the \lya luminosity distribution with respect to the general LAE population. As the sky field where observations of our proto-cluster have been performed is rather small and essentially dominated by the presence of the PCF structures, we do not consider the option of constructing a sample of low-density LAEs representative of the field population as viable. 

\section{The correlation between \lya luminosity, EW and UV continuum luminosity}
\label{ciardulloplots}

In our analyses thus far, we have assumed that there is no statistical correlation between Ly$\alpha$ equivalent width and line luminosity. By doing so, we have obtained results that are consistent with existing studies in the literature. However, if an intrinsic correlation exists between a galaxy's UV luminosity ($M_{\rm UV}$), line luminosity ($L_{{\rm Ly}\alpha}$), and EW ($W_0$), it can substantially bias the best-fit LF and EWF parameters. 

Indeed, \citet{Ando06} noted a significant deficiency of high-EW sources in UV-luminous LBGs at $z\sim 4-5$. Subsequent studies found a similar trend not only for LBGs \citep{Vanzella09,oyarzun17} but also for LAEs \citep[e.g.,][]{Shioya09}; possible causes include changing dust geometry, metallicity, and population age with galaxy's UV luminosity (and with SFR), which  alter the production rate of Ly$\alpha$ photons or modulate the likelihood of their escape through the interstellar medium \citep[e.g.,][]{Kobayashi10}. 

Given that the ``Ando effect'' is statistical in nature, a rigorous examination of such a trend, or the lack of one, requires a large sample of galaxies covering a sufficiently wide dynamic range in \lya and UV-continuum luminosity. \citet{Nilsson2009b} conducted a series of Monte Carlo realizations in which different scaling laws between the two were used to populate the galaxies according to their luminosity function. By comparing their result with the distribution of 232 LAEs and 128 LBGs at $z\sim3$ in the EW-$L_{{\rm Ly}\alpha}$ space, they concluded that no correlation is necessary to explain the observed distribution of galaxy luminosities and EWs. Furthermore, they argued that the lack of sources with high-EW and high-$L_{\rm UV}$ is simply due to them being rare objects at the extreme end of both line and continuum luminosities. Based on their study of 130 LAEs at $z\sim3$, \citet{Ciardullo2012} stated that photometric scatter plays a crucial role in creating an artificial correlation between $M_{\rm UV}$ and EW because they are anti-correlated, with EW decreasing with increasing $M_{\rm UV}$ (when other galaxy properties are fixed).

Having measured the relevant quantities (EW, $L_{{\rm Ly}\alpha}$, and $M_{\rm UV}$) of our 171 LAEs and having characterized how photometric scatter affects these measurements through our simulations (Sections~\ref{lyalewmeasurement} and \ref{simdata}), we revisit the possibility of any intrinsic correlation between $M_{\rm UV}$ and $W_0$. 

\begin{figure}
\includegraphics[width = \linewidth]{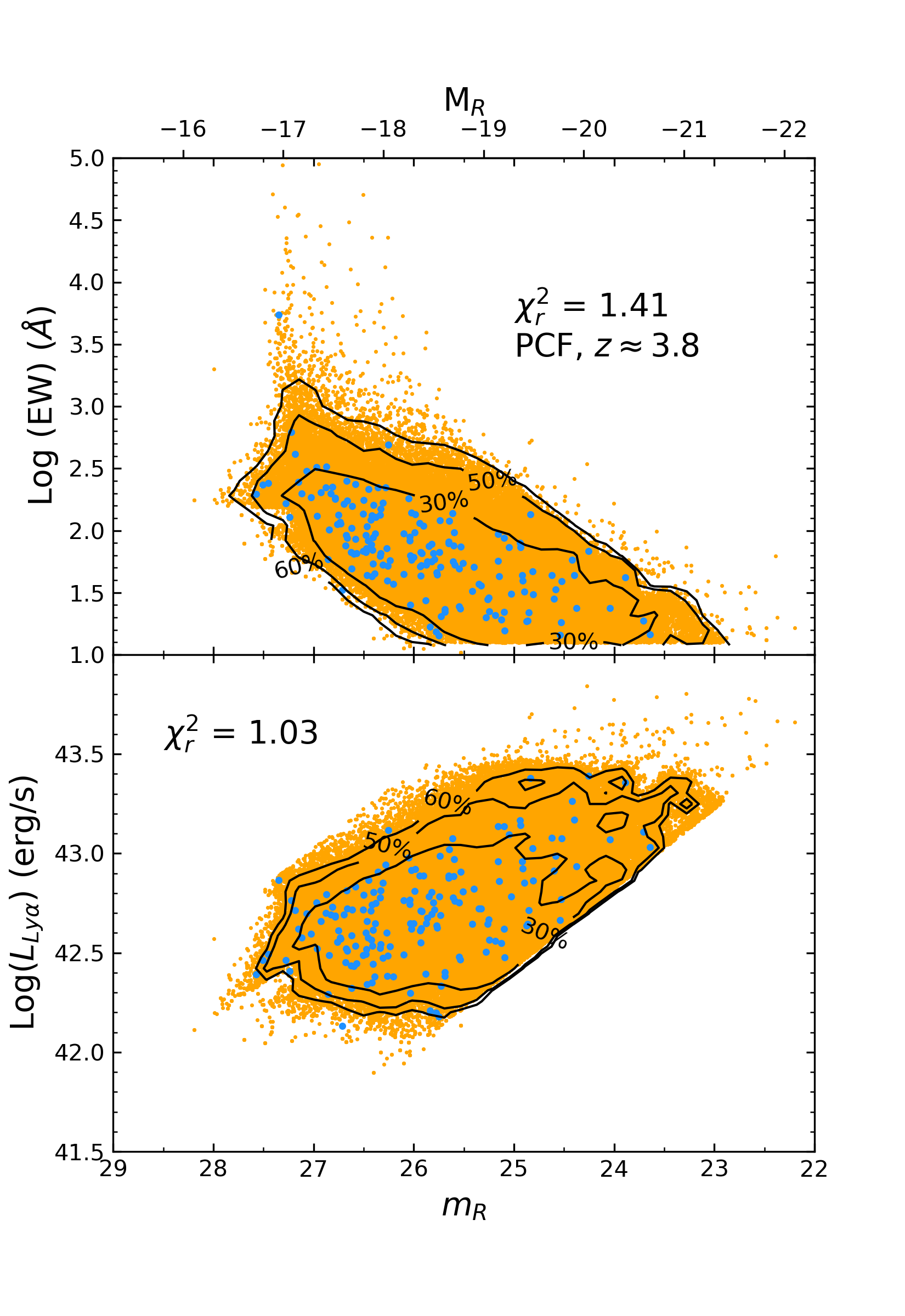}
\caption{Distribution of real and simulated LAEs on the planes defined by observable quantities. \emph{Top}: UV continuum luminosity-EW plane. \emph{Bottom}: UV continuum luminosity-\lya luminosity plane. In both panels, cyan points represent real LAEs from our PCF ($z \sim 3.790$) sample, orange points represent random extractions from our simulated LAE sample. Black contours encompass the given percentage of the simulated LAE distribution. The agreement between the simulations and real data suggests that the observed LAE sample is consistent with the \lya luminosity and EW distributions being independent of UV continuum magnitude.}
\label{ciardullopcf}
\end{figure}

We begin by evaluating the agreement between the observed and simulated data in the case of no correlation in order to establish the baseline. From our simulated LAE catalog (Section~\ref{simdata}), we randomly extract sources matching the number of real LAEs. For each simulated galaxy (with values $L_{{\rm Ly}\alpha, j}$, $W_{0,j}$), the product $\Phi_{\rm exp}(L_{{\rm Ly}\alpha, j}) \cdot \Psi_{\rm exp}(W_{0,j})$ determines the likelihood of extraction, where we assume the best-fit $\Phi(L_{{\rm Ly}\alpha})$ and $\Psi (W_0)$  for the full sample as given in Table~\ref{bestfitparams}; the procedure ensures that a mock galaxy sample is overall characterized by the identical number counts and color distributions as the real LAE sample within Poisson uncertainties. The extraction is repeated 1,000 times. In Figure~\ref{ciardullopcf}, we show the positions of real LAEs (blue) and the distribution of simulated sources (orange dots, black contour lines enclose given percentages of all simulated galaxies). It is evident that the observed galaxy distribution is  similar to the simulated one.

We bin both real and simulated data using a binsize of $\Delta M_{\rm UV} = 0.47$ mag, $\Delta W_0 = 0.2$ \AA, and $\Delta \log{L_{{\rm Ly}\alpha}} = 0.15$ for PCF. The simulated counts are normalized to match the real data, and thus can be considered as the expected value.  The reduced chi-square $\chi_r^2$ (indicated in Figure~\ref{ciardullopcf}) is close to unity in all cases, confirming our visual impression of a good agreement. 

In both $M_{\rm UV}$-$W_0$ and $M_{\rm UV}$-$L_{\mathrm{Ly}\alpha}$ planes, the two parameters are clearly correlated in a similar manner to what shown in Figure~6 of \citet{Ciardullo2012}; the lack of UV-luminous high-EW sources is also apparent.  However, the same trend seen in mock galaxies suggests that these correlations are not physical but are rather driven by the LF, EWF, LAE selection, and photometric scatter.

Our base model provides an excellent fit ($\chi_r^2 \approx 1$) to the data, suggesting that any correlation must be weak at best. Nevertheless, we investigate if a better or an equally good fit can be obtained when we require that galaxies avoid a certain zone in the parameter space. As introducing a real correlation between \lya luminosity and UV luminosity would require us to drastically change our simulation setup, we instead opted to obtain it artificially through our random extractions. Specifically, we assume that, at a fixed $m_{1700}$, the maximum or minimum $L_{\mathrm{Ly}\alpha}$ allowed for a galaxy is given by: $L_{\mathrm{Ly}\alpha} = p\cdot m_{1700} + q$. For a given set of $(p,q)$, two models can exist in which the relevant properties of galaxies can lie above or below the line, which we will refer to as $(p,q,0)$ and $(p,q,1)$, respectively. We vary the parameter range in $p = [-5,5]$ and we derive $q$ by forcing the lines to pass through the same point making this essentially a one-parameter model. The Ando effect can be emulated by a family of models with $(p,q,1)$.

\begin{figure}
\includegraphics[width = \linewidth]{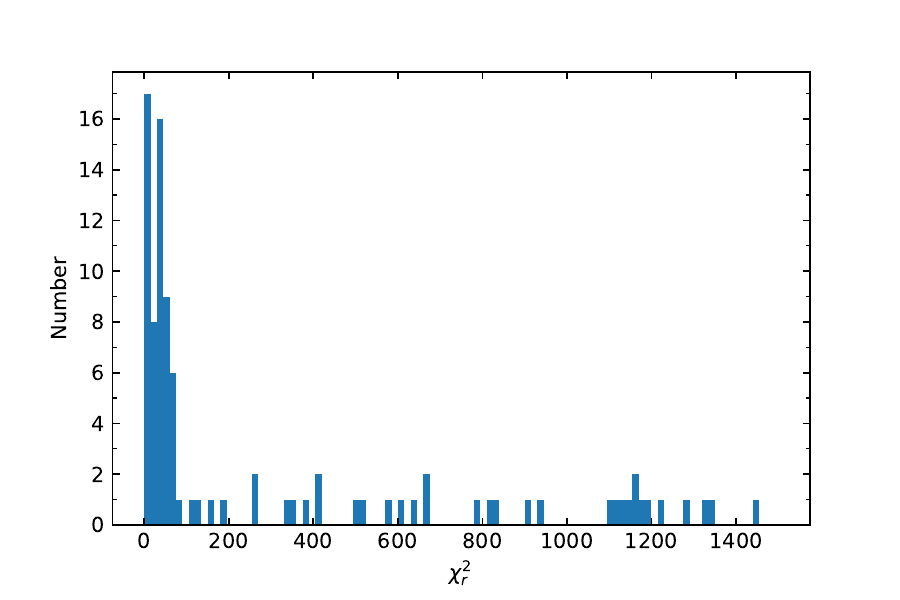}
\includegraphics[width = \linewidth]{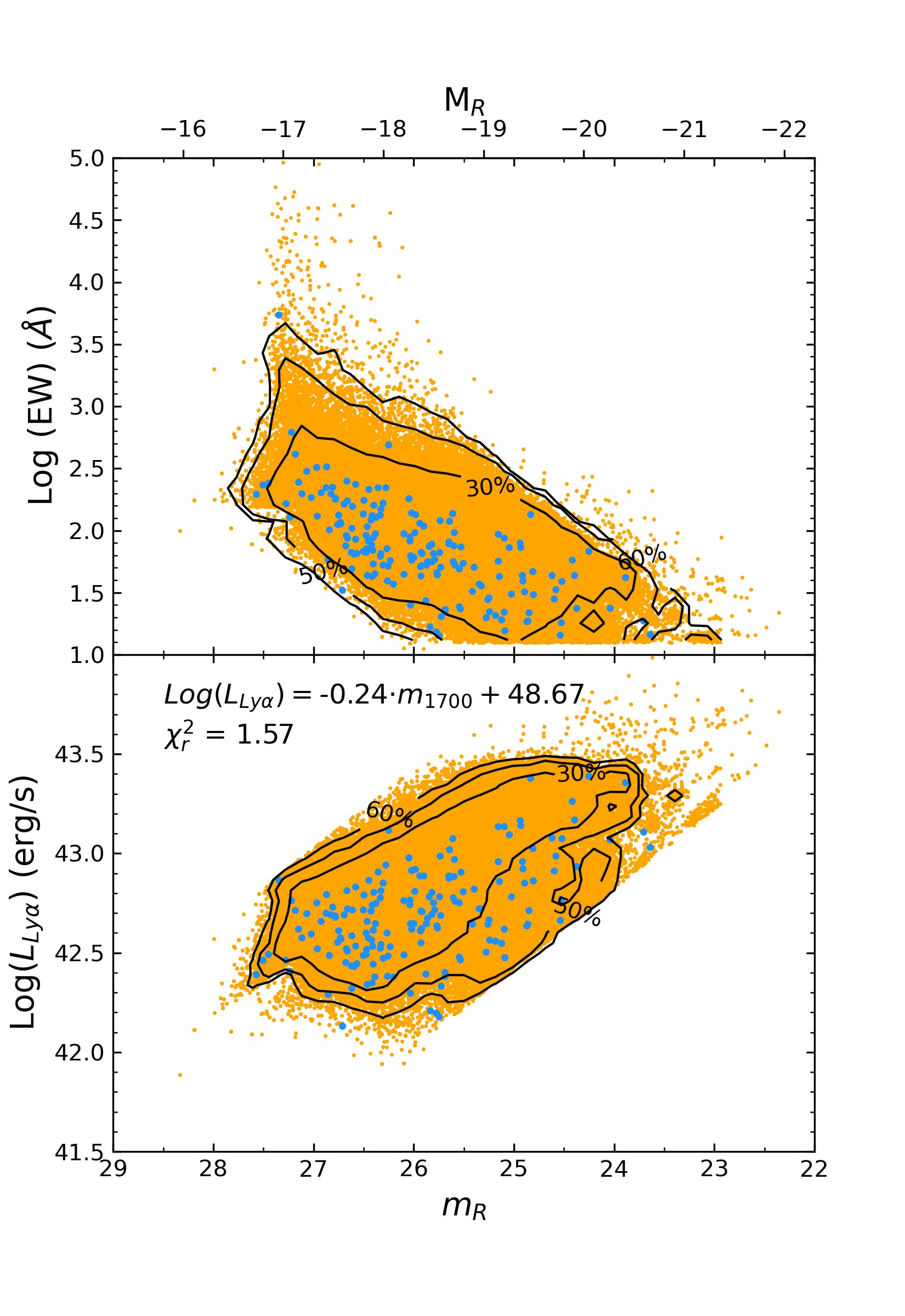}
\caption{Distribution of real and simulated LAEs on the planes defined by observable quantities, best-fit solution. \emph{Top}: the distribution of reduced $\chi^{2}_{r}$ values. \emph{Bottom}: UV continuum luminosity-EW plane and UV continuum luminosity-\lya luminosity plane. Cyan points represent real LAEs from our PCF ($z \sim 3.790$) sample, orange points represent random extractions from our simulated LAE sample. Black contours encompass the given percentages of the simulated LAE distribution. The introduction of any intrinsic correlation yields a poorer fit to the observations.}
\label{ciardullobestfit}
\end{figure}

We repeat the same extraction procedure but this time only choosing sources that satisfy the new restriction, and create the expected galaxy distribution.  In Figure \ref{ciardullobestfit}, we show the best-fit model, which is $p = -0.24_{-0.14}^{+0.07}$ and $q = 48.67_{+3.24}^{-1.62}$ \AA$\:$ for our $z \sim 3.8$ sample\footnote{The errors on the fit parameter are computed as the minimum and maximum value $p$ takes within a $1\sigma$ range of $\chi_{r}^{2}$ values centered on the best fit $p$ value. The errors on $q$ are computed by taking the $q$ value corresponding to the maximum and minimum value of $p$ as this is essentially a one-parameter fit. Notice that an increase in $p$ corresponds to a decrease in $q$ and vice versa.}, yielding a $\chi^2_r$ of 1.57, considerably worse than the best-fit with no restriction. This best-fit model is in the configuration $(p,q,0)$. In the same figure, we also show the chi-square distribution of all tested models (top panel). This distribution shows how there is a large number of solutions which yield a poor fit, extending to high chi-square values.

\begin{figure}
\includegraphics[width=\linewidth]{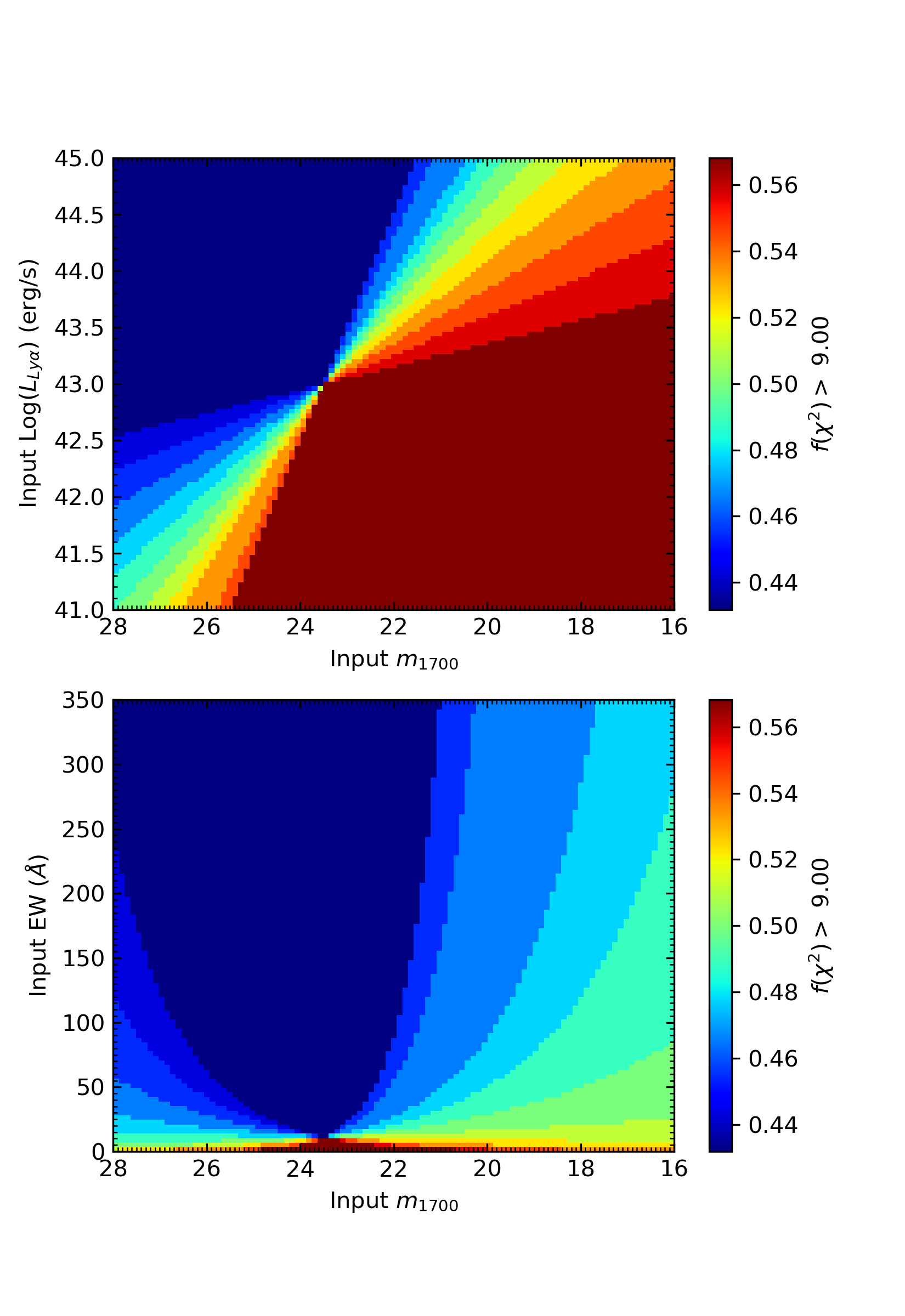}
\caption{Regions of the UV continuum-EW plane (bottom) and of the UV continuum-\lya luminosity plane (top) colour coded according to the fraction of solutions which have a reduced $\chi^{2}_{r} > 9$.}
\label{avoidance}
\end{figure}

Using the goodness-of-fit values obtained from the series of restrictive models, we identify the parameter space with the most constraining power. In practice, we define it as the region which most often yields high reduced $\chi^{2}$ values, i.e. greater than $3\sigma$. We identify this region through the fraction of solutions ($f(\chi^{2})$) that yield a reduced $\chi^{2}$ value greater than $3\sigma = 9$. Figure \ref{avoidance} shows the UV continuum-EW and the UV continuum-\lya$\:$ luminosity plane colour-coded according to $f(\chi^{2} > 9)$. It can be seen how the regions of these planes corresponding to bright LAEs are systematically disfavoured. Gathering new data which systematically better sample the high-luminosity end of the UV luminosity function and the faint end of the \lya luminosity function should provide better insight on the possible presence of a correlation between \lya luminosity and UV-continuum luminosity.

\section{Discussion}
\label{discussion}

\subsection{Implications for Galaxy Formation in proto-cluster Environment}
In this work, we have examined how the Ly$\alpha$ properties vary as a function of large-scale environment, finding no significant difference in the shape of the \lya LF and EWF between overdense regions and the general proto-cluster environment (aside from an obvious increase in the LF normalisation in the denser environments). Though limited by small number statistics, doing so allows us to contemplate on the future prospect of Ly$\alpha$ mapping as a way to study galaxies in dense proto-cluster environment, or alternatively,  to explore how to use Ly$\alpha$-emitting galaxies to search for massive proto-clusters. Such insights will serve well in the era of the Vera C. Rubin Observatory Legacy Survey of Space and Time and Hobby-Eberly Telescope Dark Energy eXperiment \citep{hill08} to develop realistic expectations. 

The present study strongly suggests that the line properties (both LF and EWF) measured for the proto-cluster LAEs are broadly consistent with that determined in the average field, in good agreement with our previous studies based on multiple proto-clusters  \citep{Dey2016,Shi19b}. This is however in mild disagreement with other works \citep[see, e.g.,][and references therein]{Lemaux2018} which found a reduced \lya emission and smaller EWs for galaxies in the protocluster environment (although, in the case of \citealt{Lemaux2018} this could be due to their galaxies being more luminous and evolved sources than what investigated here). \citet[][]{Xue2017} showed that the similarity in the line properties between proto-cluster and field LAEs extends to the circumgalactic scale \citep[but see][]{matsuda12}.

To date, only a handful of high-redshift proto-clusters have received deep wide-field  imaging with a narrow-band filter sampling their redshifted Ly$\alpha$ emission \citep{matsuda05,Lee2014,Dey2016,Xue2017,Badescu2017, Ouchi2018silverrush, Higuchi2019, Harikane2019}. Despite their small number, these case studies have demonstrated that LAEs can effectively map the large-scale structure in and around these proto-clusters showing spatial morphologies reminiscent of dark matter structures \citep[e.g.,][]{boylankolchin09}.

Taken together, these recent results appear to reinforce the notion that a deep and wide-field LAE survey will be effective in finding galaxies above a given threshold of SFR with little to no bias in regards to the large-scale environment. In the era of wide-field imaging surveys such as the Vera C. Rubin Observatory Legacy Survey of Space and Time, a sensitive wide-field ($\gtrsim 50$~deg$^2$) LAE survey of suitable depths will provide a competitive and relatively economical avenue to discover a large sample of proto-clusters.

\subsection{Redshift Evolution of the Ly$\alpha$ Emitters}
The results we obtained at $z = 3.78$ are consistent with recent studies in the literature performed at several different redshifts. \citet{Sobral2018} compiled large samples of LAEs at $z\sim 2-6$ and reported a continuous rise of $L^*$ with redshift by a factor of $\sim5$ \citep[see also][]{Konno2016}. During this period, the $\phi^*$ parameter decreases with redshift by a factor of 7, most of which occurs at $z\gtrsim 3$.  

\begin{figure*}
\centering
\includegraphics[width = 5in]{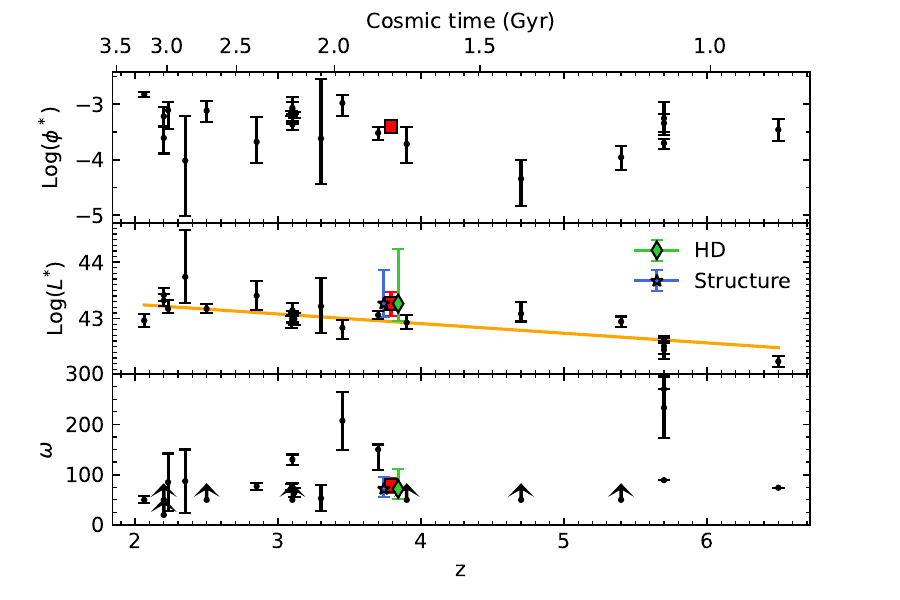}\\
\caption{Redshift evolution of the best-fit parameters for the \lya LF and EW distribution. From top to bottom panel the redshift evolution of the \citet{Schechter1976} function parameters ($\Phi^{\ast}$, $L^{\ast}$) and the exponential characteristic EW ($\omega$) is shown. In each panel, black circles are values from the literature, sorted by redshift, the red square corresponds to our high-redshift LAE sample. Values from the literature come from \citet[][$z = 2.2$]{Konno2016}, \citet[][$z = 3.1, 3.7, 5.7$]{Ouchi2008}, \citet[][$z = 3.095$]{Gronwall2007}, \citet[][$z = 3.113$]{Ciardullo2012}, \citet[][$z = 2.063$]{Guaita2010}, \citet[][$z = 2.85, 2.35, 3.3$]{Blanc2011}, \citet[][$z = 2.23$]{Sobral2017}, \citet[][$z = 3.45$]{vanBreukelen2005}, \citet[][$z = 5.7$]{Shimasaku2006}, \citet[][$z = 5.7, 6.5$]{Kashikawa2011}, \citet[][$z = 2.2, 2.5, 3.1, 3.9, 4.7, 5.4$]{Sobral2018}, and \citet[][$z = 3.1$]{Guo2020}. These literature data are corrected for different cosmology whenever applicable. In the middle panel, the orange line shows our best fit to the $\log L^{\ast} - z$ relation. In the middle and bottom panel, the green diamond and blue star show the best-fit parameters obtained for our ``HD'' and ``Structure'' samples high-density LAEs, respectively.}
\label{zevol}
\end{figure*}

Figure \ref{zevol} shows the LF and EWF best-fit parameter values from the literature sorted by increasing redshift and also includes the measurements obtained in this work. This figure shows that our estimate is within the general trend with redshift as described by the literature and that it is consistent with measurements of the LF and EWF performed at similar redshifts as our LAE sample. In particular, the $L^{\ast}$ parameter slowly decreases with increasing redshift. Our measurement is in agreement (within the errors) with values at similar redshift. We tested the possibility of fitting a linear evolution of $L^{\ast}$ with redshift and obtained a best-fit solution of 

\begin{equation}
\log L^{\ast} = -(0.17 \pm 0.02) \cdot z + (43.6 \pm 0.1)
\end{equation}

The fact that the characteristic luminosity $L^{\ast}$ has had a shallow evolution over a broad redshift range may be an indication that the properties of the LAE population have not changed dramatically across this period of cosmic time.

The LF normalization and the characteristic EW seem to have a more unclear trend (although this could be due to cosmic variance or, in the case of characteristic EW, to the different parametrisation by different works). In particular, the normalisation of the LF may be a more difficult parameter to constrain which presents more variance among the samples (e.g., several of the works mentioned in Figure \ref{zevol} rely on samples drawn from fields where structures may be present). 

The same Figure also reports our measurements of $L^{\ast}$ and characteristic EW for the LAEs identified in different environments, showing no detectable difference with the general case. Based on this analysis it can be concluded that from the viewpoint of \lya properties there seems to be no difference between LAEs in the most overdense regions and in the general proto-cluster environment (at least at a level that can be detected with our current sample size).

\subsection{Lyman-$\alpha$ escape fraction}
The current description of the physical properties of LAEs is that of dusty, low-mass, star-forming galaxies. Star-forming regions in these galaxies are responsible for the production of both the \lya radiation and the UV-continuum emission. Out of these two quantities, \lya emission is the most affected by dust extinction, while UV-continuum emission can be considered as a better tracer of the total star-formation activity, including both the obscured and the unobscured one. On the other hand, any measurement of the Star Formation Rate (SFR) derived from \lya radiation will trace only the unobscured star formation activity. For this reason, the ratio of the \lya luminosity density and of the UV luminosity density for a population of LAEs is a proxy for the fraction of unobscured star formation characterising a galaxy population and can yield information on its typical dust properties.

For this reason, we have derived the ratio of the \lya and UV luminosity densities (converted to SFRD) using the \lya LF derived in our work and UV LFs from the literature. Using the best-fit parameters for the \lya LF derived in this work, we have integrated the \citet{Schechter1976} function down to a luminosity of $1.75 \times 10^{41}~\mathrm{erg}~\mathrm{s}^{-1}$ (i.e., down to $0.04L^{\ast}_{z=3}$ adopting the value from \citealt{Gronwall2007}) to be consistent with what done by \citet{Sobral2018}.

We use the best-fit parameters for the UV LF from Table 6 of \citet[][at $z \approx 3.8$ consistent with our $z \approx 3.790$ sample]{Bouwens2015} together with the mean dust extinction correction factor of 2.4 quoted in the text for the relevant redshift. We integrate the UV luminosity functions down to $0.04L^{\ast}_{z=3}$ as derived by \citet{Steidel1999}.

Following Equation~11 of \citet{Sobral2018}, we derive the ratio of the \lya-to-UV SFRD as:
\begin{equation}
\frac{\mathrm{SFRD}_{\mathrm{Ly}{\alpha}}}{\mathrm{SFRD}_{\mathrm{UV}}} = \xi_{\rm ion, N} \cdot f_{\rm esc}
\end{equation}
where $\xi_{\rm ion, N} = \xi_{\rm ion}/(1.3 \times 10^{25}~ \mathrm{Hz}~ \mathrm{erg}^{-1})$. Following equation 9 of \citet{Sobral2018}. 
\begin{equation}
\xi_{\rm ion} \cdot f_{\rm esc} = \frac{\rho_{\mathrm{Ly}{\alpha}}}{8.7c_{\mathrm{H}\alpha}\rho_{\mathrm{UV}}}
\end{equation}
where $\xi_{\rm ion}$ is the production efficiency of Hydrogen ionizing photons and $c_{\mathrm{H}\alpha} = 1.37 \times 10^{12} \mathrm{erg}$ is the recombination coefficient.

\begin{figure}
\centering
\includegraphics[width = \linewidth]{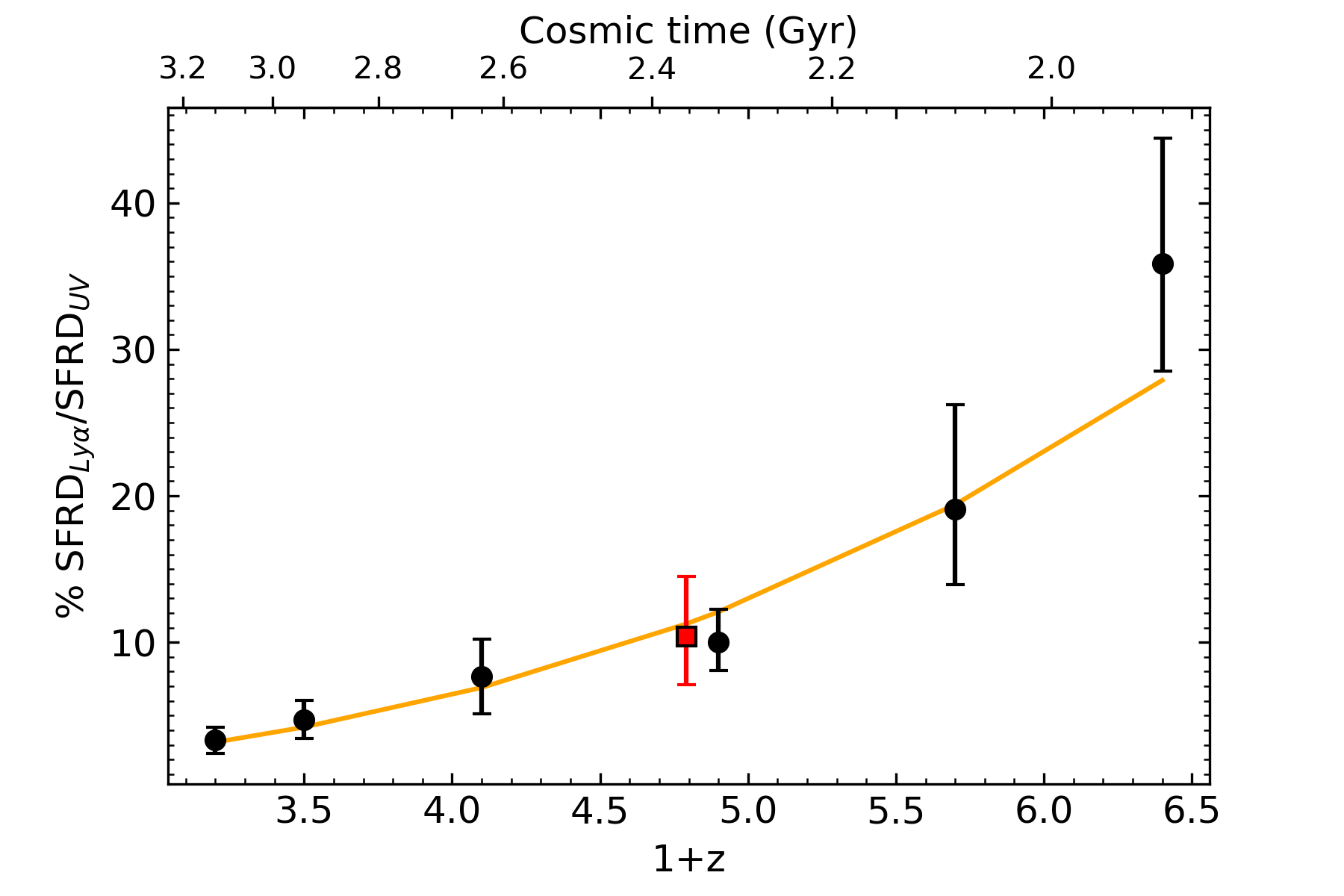}\\
\caption{Redshift evolution of the ratio of the SFRD derived from \lya radiation and from the UV. The red square corresponds to our high-redshift LAE sample. Black points are derived from Table 6 of \citet{Sobral2018} divided by the dust-corrected values from Table 7 and Table 5 of \citet{Bouwens2015} and \citet{ReddySteidel2009}, respectively.}
\label{escapefraction}
\end{figure}

Figure \ref{escapefraction} shows the evolution with redshift of $\mathrm{SFRD}_{\mathrm{Ly}{\alpha}}/\mathrm{SFRD}_{\mathrm{UV}} \propto \xi_{\rm ion} \cdot f_{\rm esc}$ for our work, together with values from \citet{Sobral2018}. The overall trend is for the SFRD ratio to slowly increase with redshift from below $5\%$ to above $\sim 10\%$ in the redshift range $z \sim 2-4$, reaching $35\%$ at redshift $z \sim 5.5$. Our estimated ratio of ${\rm SFRD_{\rm Ly\alpha}/SFRD_{\rm UV}}$ is consistent with other measurements from the literature at a similar redshift, with a value of $\sim 10\%$. We fit the redshift evolution of $\mathrm{SFRD}_{\mathrm{Ly}{\alpha}}/\mathrm{SFRD}_{\mathrm{UV}}$ with an exponential function of $1+z$ and obtain a best-fit result in the form of

\begin{equation}
\frac{\mathrm{SFRD}_{\mathrm{Ly}{\alpha}}}{\mathrm{SFRD}_{\mathrm{UV}}} = 8.3 \times 10^{-4} \cdot (1+z)^{3.134}
\end{equation} 

An increase in the SFRD with redshift corresponds to both a change in the amount of ionizing radiation available, as well as a change in the ISM properties. In particular, \citet{Sobral2018} argue that modelling the change in $\xi_{\rm ion}$ as a linear evolution with $(1+z)$ results in substantial evolution of $f_{\rm esc}$ with redshift. The general trend seems to suggest that with cosmic time a larger fraction of SFR activity happens in dust-obscured environments. At $z\approx3.8$, the majority of stars are formed in a dust-rich environment, allowing for a very small amount of \lya radiation to escape from these galaxies. Several high-redshift proto-clusters show evidence of massive dusty galaxies; e.g., dusty sub-millimeter galaxies are detected in a $z = 4.3$ proto-cluster \citep{Miller2018, Hill2020, Rotermund2021}. The detection of an increase in the escape fraction of LAEs with redshift suggests the scenario in which the amount of dust in this lower mass galaxies increases with cosmic time.

\section{Conclusions and summary}
\label{conclusions}
In this work, we present the \lya luminosity function and equivalent width distribution of a sample of LAEs at $z \approx 3.8$ in the field of a dense proto-cluster. The field contains one of the largest and most massive proto-clusters known to date with more than 60 spectroscopically confirmed members, providing a rare opportunity to examine how these properties change with the large-scale environment of galaxies. Based on the analysis of this sample, we find: 

\begin{enumerate}

\item In terms of normalisation, the LF of our entire LAE sample is a factor of $\sim 2.0-2.5$ higher than that in the average field \citep[e.g.][]{Ouchi2008, Cassata2011} while the overall shape remains similar, reflecting the significant galaxy overdensity in the proto-cluster region.

\item There is no significant difference in the \lya luminosity function and equivalent width distributions of the LAE population inhabiting the most overdense regions with respect to those of the LAE population inhabiting the general proto-cluster environment in the PCF field (other than the overall normalization). From the \lya viewpoint, the implication is that LAEs in the dense proto-cluster environment form in a manner undistinguishable from that in the average field. However, our conclusion may be tempered by the reduced size of our sample.

\item The measured \lya luminosity function at $z \approx 3.8$ is broadly in agreement with the existing measures in the literature. The characteristic luminosity $L^{\ast}$ shows a mild evolution, decreasing with increasing redshift. We find that the EW distribution does not evolve with redshift, in agreement with the existing measures. 

\item By measuring the ratio of the integrated \lya LF to the integrated UV LF from the literature we are able to estimate the ratio of the unobscured-to-total star-formation rate density. This is a proxy for the escape fraction of our LAE sample. Our measurement of $\xi_{ion} \cdot f_{esc} \sim 10\%$ is consistent with values from the literature. We find the escape fraction to increase with redshift, with a very clearly defined trend.
 
\item Using a large suite of image simulations, we examine the possibility that an intrinsic correlation exists between a galaxy's UV luminosity, line luminosity, and EW. Such correlations would introduce a bias in the measurement of the \lya luminosity function and EWF parameters. We find that the apparent correlation in our measurements can be explained by photometric selection effects. Introduction of any intrinsic correlation yields a poorer fit to the observations.
\end{enumerate}

\section*{Acknowledgements}
This research has been supported by the funding for the ByoPiC project from the European Research Council (ERC) under the European Union's Horizon 2020 research and innovation programme grant agreement ERC-2015-AdG 695561. Based in part on observations at Kitt Peak National Observatory at NSF's NOIRLab (NOIRLab Prop. ID 2012A-0454, 2014A-0164; PI: Kyoung-Soo Lee), which is managed by the Association of Universities for Research in Astronomy (AURA) under a cooperative agreement with the National Science Foundation. The authors are honored to be permitted to conduct astronomical research on Iolkam Du'ag (Kitt Peak), a mountain with particular significance to the Tohono O'odham. This work also made use of images and data products provided by the NOAO Deep Wide-Field Survey (Jannuzi and Dey 1999), which was supported by the National Optical Astronomy Observatory (NOAO, now NOIRLab). The work of A. Dey is supported by NOIRLab, which is managed by the Association of Universities for Research in Astronomy (AURA) under a cooperative agreement with the National Science Foundation. The National Radio Astronomy Observatory is a facility of the National Science Foundation operated under cooperative agreement by Associated Universities, Inc.

\bibliography{Lya_lf_bib}{}

\begin{thebibliography}{}
\expandafter\ifx\csname natexlab\endcsname\relax\def\natexlab#1{#1}\fi
\providecommand{\url}[1]{\href{#1}{#1}}
\providecommand{\dodoi}[1]{doi:~\href{http://doi.org/#1}{\nolinkurl{#1}}}
\providecommand{\doeprint}[1]{\href{http://ascl.net/#1}{\nolinkurl{http://ascl.net/#1}}}
\providecommand{\doarXiv}[1]{\href{https://arxiv.org/abs/#1}{\nolinkurl{https://arxiv.org/abs/#1}}}

\bibitem[{{Adams} {et~al.}(2009){Adams}, {Hill}, \& {MacQueen}}]{Adams2009}
{Adams}, J.~J., {Hill}, G.~J., \& {MacQueen}, P.~J. 2009, \apj, 694, 314,
  \dodoi{10.1088/0004-637X/694/1/314}

\bibitem[{Adams {et~al.}(2011)Adams, Blanc, Hill, Gebhardt, Drory, Hao, Bender,
  Byun, Ciardullo, Cornell, Finkelstein, Fry, Gawiser, Gronwall, Hopp, Jeong,
  Kelz, Kelzenberg, Komatsu, MacQueen, Murphy, Odoms, Roth, Schneider, Tufts,
  \& Wilkinson}]{Adams2011}
Adams, J.~J., Blanc, G.~A., Hill, G.~J., {et~al.} 2011, The Astrophysical
  Journal Supplement Series, 192, 5.
\newblock \url{http://stacks.iop.org/0067-0049/192/i=1/a=5}

\bibitem[{Ajiki {et~al.}(2006)Ajiki, Mobasher, Taniguchi, Shioya, Nagao,
  Murayama, \& Sasaki}]{Ajiki2006}
Ajiki, M., Mobasher, B., Taniguchi, Y., {et~al.} 2006, The Astrophysical
  Journal, 638, 596.
\newblock \url{http://stacks.iop.org/0004-637X/638/i=2/a=596}

\bibitem[{{Ando} {et~al.}(2006){Ando}, {Ohta}, {Iwata}, {Akiyama}, {Aoki}, \&
  {Tamura}}]{Ando06}
{Ando}, M., {Ohta}, K., {Iwata}, I., {et~al.} 2006, \apjl, 645, L9,
  \dodoi{10.1086/505652}

\bibitem[{{Blanc} {et~al.}(2011){Blanc}, {Adams}, {Gebhardt}, {Hill}, {Drory},
  {Hao}, {Bender}, {Ciardullo}, {Finkelstein}, {Fry}, {Gawiser}, {Gronwall},
  {Hopp}, {Jeong}, {Kelzenberg}, {Komatsu}, {MacQueen}, {Murphy}, {Roth},
  {Schneider}, \& {Tufts}}]{Blanc2011}
{Blanc}, G.~A., {Adams}, J.~J., {Gebhardt}, K., {et~al.} 2011, \apj, 736, 31,
  \dodoi{10.1088/0004-637X/736/1/31}

\bibitem[{{Bouwens} {et~al.}(2004){Bouwens}, {Illingworth}, {Blakeslee},
  {Broadhurst}, \& {Franx}}]{Bouwens2004}
{Bouwens}, R.~J., {Illingworth}, G.~D., {Blakeslee}, J.~P., {Broadhurst},
  T.~J., \& {Franx}, M. 2004, \apj, 611, L1, \dodoi{10.1086/423786}

\bibitem[{{Bouwens} {et~al.}(2007){Bouwens}, {Illingworth}, {Franx}, \&
  {Ford}}]{Bouwens2007}
{Bouwens}, R.~J., {Illingworth}, G.~D., {Franx}, M., \& {Ford}, H. 2007, \apj,
  670, 928, \dodoi{10.1086/521811}

\bibitem[{{Bouwens} {et~al.}(2015){Bouwens}, {Illingworth}, {Oesch}, {Trenti},
  {Labb{\'e}}, {Bradley}, {Carollo}, {van Dokkum}, {Gonzalez}, {Holwerda},
  {Franx}, {Spitler}, {Smit}, \& {Magee}}]{Bouwens2015}
{Bouwens}, R.~J., {Illingworth}, G.~D., {Oesch}, P.~A., {et~al.} 2015, \apj,
  803, 34, \dodoi{10.1088/0004-637X/803/1/34}

\bibitem[{{Boylan-Kolchin} {et~al.}(2009){Boylan-Kolchin}, {Springel}, {White},
  {Jenkins}, \& {Lemson}}]{boylankolchin09}
{Boylan-Kolchin}, M., {Springel}, V., {White}, S.~D.~M., {Jenkins}, A., \&
  {Lemson}, G. 2009, \mnras, 398, 1150,
  \dodoi{10.1111/j.1365-2966.2009.15191.x}

\bibitem[{{Bruzual} \& {Charlot}(2003)}]{BC03}
{Bruzual}, G., \& {Charlot}, S. 2003, \mnras, 344, 1000,
  \dodoi{10.1046/j.1365-8711.2003.06897.x}

\bibitem[{{B{\u{a}}descu} {et~al.}(2017){B{\u{a}}descu}, {Yang}, {Bertoldi},
  {Zabludoff}, {Karim}, \& {Magnelli}}]{Badescu2017}
{B{\u{a}}descu}, T., {Yang}, Y., {Bertoldi}, F., {et~al.} 2017, \apj, 845, 172,
  \dodoi{10.3847/1538-4357/aa8220}

\bibitem[{{Cassata} {et~al.}(2011){Cassata}, {Le F{\`e}vre}, {Garilli},
  {Maccagni}, {Le Brun}, {Scodeggio}, {Tresse}, {Ilbert}, {Zamorani},
  {Cucciati}, {Contini}, {Bielby}, {Mellier}, {McCracken}, {Pollo},
  {Zanichelli}, {Bardelli}, {Cappi}, {Pozzetti}, {Vergani}, \&
  {Zucca}}]{Cassata2011}
{Cassata}, P., {Le F{\`e}vre}, O., {Garilli}, B., {et~al.} 2011, \aap, 525,
  A143, \dodoi{10.1051/0004-6361/201014410}

\bibitem[{{Charlot} \& {Fall}(1993)}]{CharlotFall1993}
{Charlot}, S., \& {Fall}, S.~M. 1993, \apj, 415, 580, \dodoi{10.1086/173187}

\bibitem[{{Ciardullo} {et~al.}(2012){Ciardullo}, {Gronwall}, {Wolf},
  {McCathran}, {Bond}, {Gawiser}, {Guaita}, {Feldmeier}, {Treister}, {Padilla},
  {Francke}, {Matkovi{\'c}}, {Altmann}, \& {Herrera}}]{Ciardullo2012}
{Ciardullo}, R., {Gronwall}, C., {Wolf}, C., {et~al.} 2012, \apj, 744, 110,
  \dodoi{10.1088/0004-637X/744/2/110}

\bibitem[{{Cowie} \& {Hu}(1998)}]{CowieHu1998}
{Cowie}, L.~L., \& {Hu}, E.~M. 1998, \aj, 115, 1319, \dodoi{10.1086/300309}

\bibitem[{{Dey} {et~al.}(2016){Dey}, {Lee}, {Reddy}, {Cooper}, {Inami}, {Hong},
  {Gonzalez}, \& {Jannuzi}}]{Dey2016}
{Dey}, A., {Lee}, K.-S., {Reddy}, N., {et~al.} 2016, \apj, 823, 11,
  \dodoi{10.3847/0004-637X/823/1/11}

\bibitem[{{Dijkstra} {et~al.}(2006){Dijkstra}, {Haiman}, \&
  {Spaans}}]{Dijkstra2006}
{Dijkstra}, M., {Haiman}, Z., \& {Spaans}, M. 2006, \apj, 649, 37,
  \dodoi{10.1086/506244}

\bibitem[{{Dijkstra} {et~al.}(2007){Dijkstra}, {Lidz}, \&
  {Wyithe}}]{Dijkstra2007}
{Dijkstra}, M., {Lidz}, A., \& {Wyithe}, J. S.~B. 2007, \mnras, 377, 1175,
  \dodoi{10.1111/j.1365-2966.2007.11666.x}

\bibitem[{{Faber} {et~al.}(2003){Faber}, {Phillips}, {Kibrick}, {Alcott},
  {Allen}, {Burrous}, {Cantrall}, {Clarke}, {Coil}, {Cowley}, {Davis}, {Deich},
  {Dietsch}, {Gilmore}, {Harper}, {Hilyard}, {Lewis}, {McVeigh}, {Newman},
  {Osborne}, {Schiavon}, {Stover}, {Tucker}, {Wallace}, {Wei}, {Wirth}, \&
  {Wright}}]{Faber2003}
{Faber}, S.~M., {Phillips}, A.~C., {Kibrick}, R.~I., {et~al.} 2003, in
  \procspie, Vol. 4841, Instrument Design and Performance for Optical/Infrared
  Ground-based Telescopes, ed. M.~{Iye} \& A.~F.~M. {Moorwood}, 1657--1669,
  \dodoi{10.1117/12.460346}

\bibitem[{{Ferguson} {et~al.}(2004){Ferguson}, {Dickinson}, {Giavalisco},
  {Kretchmer}, {Ravindranath}, {Idzi}, {Taylor}, {Conselice}, {Fall},
  {Gardner}, {Livio}, {Madau}, {Moustakas}, {Papovich}, {Somerville},
  {Spinrad}, \& {Stern}}]{Ferguson2004}
{Ferguson}, H.~C., {Dickinson}, M., {Giavalisco}, M., {et~al.} 2004, \apjl,
  600, L107, \dodoi{10.1086/378578}

\bibitem[{{Finkelstein} {et~al.}(2009){Finkelstein}, {Rhoads}, {Malhotra}, \&
  {Grogin}}]{Finkelstein2009}
{Finkelstein}, S.~L., {Rhoads}, J.~E., {Malhotra}, S., \& {Grogin}, N. 2009,
  \apj, 691, 465, \dodoi{10.1088/0004-637X/691/1/465}

\bibitem[{{Fossati} {et~al.}(2021){Fossati}, {Fumagalli}, {Lofthouse}, {Dutta},
  {Cantalupo}, {Arrigoni Battaia}, {Fynbo}, {Lusso}, {Murphy}, {Prochaska},
  {Theuns}, \& {Cooke}}]{Fumagalli2021}
{Fossati}, M., {Fumagalli}, M., {Lofthouse}, E.~K., {et~al.} 2021, \mnras, 503,
  3044, \dodoi{10.1093/mnras/stab660}

\bibitem[{{Gawiser} {et~al.}(2006{\natexlab{a}}){Gawiser}, {van Dokkum},
  {Herrera}, {Maza}, {Castander}, {Infante}, {Lira}, {Quadri}, {Toner},
  {Treister}, {Urry}, {Altmann}, {Assef}, {Christlein}, {Coppi}, {Dur{\'a}n},
  {Franx}, {Galaz}, {Huerta}, {Liu}, {L{\'o}pez}, {M{\'e}ndez}, {Moore},
  {Rubio}, {Ruiz}, {Toft}, \& {Yi}}]{Gawiser2006MUSYC}
{Gawiser}, E., {van Dokkum}, P.~G., {Herrera}, D., {et~al.} 2006{\natexlab{a}},
  The Astrophysical Journal Supplement Series, 162, 1, \dodoi{10.1086/497644}

\bibitem[{{Gawiser} {et~al.}(2006{\natexlab{b}}){Gawiser}, {van Dokkum},
  {Gronwall}, {Ciardullo}, {Blanc}, {Castander}, {Feldmeier}, {Francke},
  {Franx}, {Haberzettl}, {Herrera}, {Hickey}, {Infante}, {Lira}, {Maza},
  {Quadri}, {Richardson}, {Schawinski}, {Schirmer}, {Taylor}, {Treister},
  {Urry}, \& {Virani}}]{Gawiser2006}
{Gawiser}, E., {van Dokkum}, P.~G., {Gronwall}, C., {et~al.}
  2006{\natexlab{b}}, \apj, 642, L13, \dodoi{10.1086/504467}

\bibitem[{{Gawiser} {et~al.}(2007){Gawiser}, {Francke}, {Lai}, {Schawinski},
  {Gronwall}, {Ciardullo}, {Quadri}, {Orsi}, {Barrientos}, {Blanc}, {Fazio},
  {Feldmeier}, {Huang}, {Infante}, {Lira}, {Padilla}, {Taylor}, {Treister},
  {Urry}, {van Dokkum}, \& {Virani}}]{Gawiser2007}
{Gawiser}, E., {Francke}, H., {Lai}, K., {et~al.} 2007, \apj, 671, 278,
  \dodoi{10.1086/522955}

\bibitem[{{Gronwall} {et~al.}(2007){Gronwall}, {Ciardullo}, {Hickey},
  {Gawiser}, {Feldmeier}, {van Dokkum}, {Urry}, {Herrera}, {Lehmer}, {Infante},
  {Orsi}, {Marchesini}, {Blanc}, {Francke}, {Lira}, \&
  {Treister}}]{Gronwall2007}
{Gronwall}, C., {Ciardullo}, R., {Hickey}, T., {et~al.} 2007, \apj, 667, 79,
  \dodoi{10.1086/520324}

\bibitem[{{Guaita} {et~al.}(2010){Guaita}, {Gawiser}, {Padilla}, {Francke},
  {Bond}, {Gronwall}, {Ciardullo}, {Feldmeier}, {Sinawa}, {Blanc}, \&
  {Virani}}]{Guaita2010}
{Guaita}, L., {Gawiser}, E., {Padilla}, N., {et~al.} 2010, \apj, 714, 255,
  \dodoi{10.1088/0004-637X/714/1/255}

\bibitem[{{Guo} {et~al.}(2020){Guo}, {Jiang}, {Egami}, {Ning}, {Zheng}, \&
  {Ho}}]{Guo2020}
{Guo}, Y., {Jiang}, L., {Egami}, E., {et~al.} 2020, \apj, 902, 137,
  \dodoi{10.3847/1538-4357/abb59a}

\bibitem[{{Hansen} \& {Oh}(2006)}]{HansenOh2006}
{Hansen}, M., \& {Oh}, S.~P. 2006, \mnras, 367, 979,
  \dodoi{10.1111/j.1365-2966.2005.09870.x}

\bibitem[{{Harikane} {et~al.}(2019){Harikane}, {Ouchi}, {Ono}, {Fujimoto},
  {Donevski}, {Shibuya}, {Faisst}, {Goto}, {Hatsukade}, {Kashikawa}, {Kohno},
  {Hashimoto}, {Higuchi}, {Inoue}, {Lin}, {Martin}, {Overzier}, {Smail},
  {Toshikawa}, {Umehata}, {Ao}, {Chapman}, {Clements}, {Im}, {Jing},
  {Kawaguchi}, {Lee}, {Lee}, {Lin}, {Matsuoka}, {Marinello}, {Nagao},
  {Onodera}, {Toft}, \& {Wang}}]{Harikane2019}
{Harikane}, Y., {Ouchi}, M., {Ono}, Y., {et~al.} 2019, \apj, 883, 142,
  \dodoi{10.3847/1538-4357/ab2cd5}

\bibitem[{{Hashimoto} {et~al.}(2013){Hashimoto}, {Ouchi}, {Shimasaku}, {Ono},
  {Nakajima}, {Rauch}, {Lee}, \& {Okamura}}]{Hashimoto2013}
{Hashimoto}, T., {Ouchi}, M., {Shimasaku}, K., {et~al.} 2013, \apj, 765,
  \dodoi{10.1088/0004-637X/765/1/70}

\bibitem[{{Hayes} {et~al.}(2010){Hayes}, {{\"O}stlin}, {Schaerer}, {Mas-Hesse},
  {Leitherer}, {Atek}, {Kunth}, {Verhamme}, {de Barros}, \&
  {Melinder}}]{Hayes2010}
{Hayes}, M., {{\"O}stlin}, G., {Schaerer}, D., {et~al.} 2010, \nat, 464, 562,
  \dodoi{10.1038/nature08881}

\bibitem[{{Herenz} {et~al.}(2019){Herenz}, {Wisotzki}, {Saust}, {Kerutt},
  {Urrutia}, {Diener}, {Schmidt}, {Marino}, {de la Vieuville}, {Boogaard},
  {Schaye}, {Guiderdoni}, {Richard}, \& {Bacon}}]{Herenz2019}
{Herenz}, E.~C., {Wisotzki}, L., {Saust}, R., {et~al.} 2019, \aap, 621, A107,
  \dodoi{10.1051/0004-6361/201834164}

\bibitem[{{Higuchi} {et~al.}(2019){Higuchi}, {Ouchi}, {Ono}, {Shibuya},
  {Toshikawa}, {Harikane}, {Kojima}, {Chiang}, {Egami}, {Kashikawa},
  {Overzier}, {Konno}, {Inoue}, {Hasegawa}, {Fujimoto}, {Goto}, {Ishikawa},
  {Ito}, {Komiyama}, \& {Tanaka}}]{Higuchi2019}
{Higuchi}, R., {Ouchi}, M., {Ono}, Y., {et~al.} 2019, \apj, 879, 28,
  \dodoi{10.3847/1538-4357/ab2192}

\bibitem[{{Hill} {et~al.}(2008{\natexlab{a}}){Hill}, {Gebhardt}, {Komatsu},
  {Drory}, {MacQueen}, {Adams}, {Blanc}, {Koehler}, {Rafal}, {Roth}, {Kelz},
  {Gronwall}, {Ciardullo}, \& {Schneider}}]{Hill2008}
{Hill}, G.~J., {Gebhardt}, K., {Komatsu}, E., {et~al.} 2008{\natexlab{a}}, in
  Astronomical Society of the Pacific Conference Series, Vol. 399, Panoramic
  Views of Galaxy Formation and Evolution, ed. T.~{Kodama}, T.~{Yamada}, \&
  K.~{Aoki}, 115.
\newblock \doarXiv{0806.0183}

\bibitem[{{Hill} {et~al.}(2008{\natexlab{b}}){Hill}, {Gebhardt}, {Komatsu},
  {Drory}, {MacQueen}, {Adams}, {Blanc}, {Koehler}, {Rafal}, {Roth}, {Kelz},
  {Gronwall}, {Ciardullo}, \& {Schneider}}]{hill08}
{Hill}, G.~J., {Gebhardt}, K., {Komatsu}, E., {et~al.} 2008{\natexlab{b}}, in
  Astronomical Society of the Pacific Conference Series, Vol. 399, Panoramic
  Views of Galaxy Formation and Evolution, ed. T.~{Kodama}, T.~{Yamada}, \&
  K.~{Aoki}, 115.
\newblock \doarXiv{0806.0183}

\bibitem[{{Hill} {et~al.}(2020){Hill}, {Chapman}, {Scott}, {Apostolovski},
  {Aravena}, {B{\'e}thermin}, {Bradford}, {Canning}, {De Breuck}, {Dong},
  {Gonzalez}, {Greve}, {Hayward}, {Hezaveh}, {Litke}, {Malkan}, {Marrone},
  {Phadke}, {Reuter}, {Rotermund}, {Spilker}, {Vieira}, \&
  {Wei{\ss}}}]{Hill2020}
{Hill}, R., {Chapman}, S., {Scott}, D., {et~al.} 2020, \mnras, 495, 3124,
  \dodoi{10.1093/mnras/staa1275}

\bibitem[{{Hu} {et~al.}(2002){Hu}, {Cowie}, {McMahon}, {Capak}, {Iwamuro},
  {Kneib}, {Maihara}, \& {Motohara}}]{Hu2002}
{Hu}, E.~M., {Cowie}, L.~L., {McMahon}, R.~G., {et~al.} 2002, \apj, 568, L75,
  \dodoi{10.1086/340424}

\bibitem[{{Hu} {et~al.}(2019){Hu}, {Wang}, {Zheng}, {Malhotra}, {Rhoads},
  {Infante}, {Barrientos}, {Yang}, {Jiang}, {Kang}, {Perez}, {Wold}, {Hibon},
  {Jiang}, {Khostovan}, {Valdes}, {Walker}, {Galaz}, {Coughlin}, {Harish},
  {Kong}, {Pharo}, \& {Zheng}}]{WeidaHu2019}
{Hu}, W., {Wang}, J., {Zheng}, Z.-Y., {et~al.} 2019, \apj, 886, 90,
  \dodoi{10.3847/1538-4357/ab4cf4}

\bibitem[{{Hu} {et~al.}(2021){Hu}, {Wang}, {Infante}, {Rhoads}, {Zheng},
  {Yang}, {Malhotra}, {Barrientos}, {Jiang}, {Gonz{\'a}lez-L{\'o}pez},
  {Prieto}, {Perez}, {Hibon}, {Galaz}, {Coughlin}, {Harish}, {Kong}, {Kang},
  {Khostovan}, {Pharo}, {Valdes}, {Wold}, {Walker}, \& {Zheng}}]{WeidaHu2021}
{Hu}, W., {Wang}, J., {Infante}, L., {et~al.} 2021, Nature Astronomy, 5, 485,
  \dodoi{10.1038/s41550-020-01291-y}

\bibitem[{{Inoue} {et~al.}(2014){Inoue}, {Shimizu}, {Iwata}, \&
  {Tanaka}}]{Inoue2014}
{Inoue}, A.~K., {Shimizu}, I., {Iwata}, I., \& {Tanaka}, M. 2014, \mnras, 442,
  1805, \dodoi{10.1093/mnras/stu936}

\bibitem[{{Ito} {et~al.}(2020){Ito}, {Kashikawa}, {Toshikawa}, {Overzier},
  {Kubo}, {Uchiyama}, {Liang}, {Onoue}, {Tanaka}, {Komiyama}, {Lee}, {Lin},
  {Marinello}, {Martin}, \& {Shibuya}}]{Ito2020}
{Ito}, K., {Kashikawa}, N., {Toshikawa}, J., {et~al.} 2020, \apj, 899, 5,
  \dodoi{10.3847/1538-4357/aba269}

\bibitem[{{Iye} {et~al.}(2006){Iye}, {Ota}, {Kashikawa}, {Furusawa},
  {Hashimoto}, {Hattori}, {Matsuda}, {Morokuma}, {Ouchi}, \&
  {Shimasaku}}]{Iye2006}
{Iye}, M., {Ota}, K., {Kashikawa}, N., {et~al.} 2006, \nat, 443, 186,
  \dodoi{10.1038/nature05104}

\bibitem[{{Jannuzi} \& {Dey}(1999)}]{JannuziDey1999}
{Jannuzi}, B.~T., \& {Dey}, A. 1999, in Astronomical Society of the Pacific
  Conference Series, Vol. 191, Photometric Redshifts and the Detection of High
  Redshift Galaxies, ed. R.~{Weymann}, L.~{Storrie-Lombardi}, M.~{Sawicki}, \&
  R.~{Brunner}, 111

\bibitem[{{Kashikawa} {et~al.}(2011){Kashikawa}, {Shimasaku}, {Matsuda},
  {Egami}, {Jiang}, {Nagao}, {Ouchi}, {Malkan}, {Hattori}, {Ota}, {Taniguchi},
  {Okamura}, {Ly}, {Iye}, {Furusawa}, {Shioya}, {Shibuya}, {Ishizaki}, \&
  {Toshikawa}}]{Kashikawa2011}
{Kashikawa}, N., {Shimasaku}, K., {Matsuda}, Y., {et~al.} 2011, \apj, 734, 119,
  \dodoi{10.1088/0004-637X/734/2/119}

\bibitem[{{Kobayashi} {et~al.}(2010){Kobayashi}, {Totani}, \&
  {Nagashima}}]{Kobayashi10}
{Kobayashi}, M.~A.~R., {Totani}, T., \& {Nagashima}, M. 2010, \apj, 708, 1119,
  \dodoi{10.1088/0004-637X/708/2/1119}

\bibitem[{{Kodaira} {et~al.}(2003){Kodaira}, {Taniguchi}, {Kashikawa}, {Kaifu},
  {Ando}, {Karoji}, {Ajiki}, {Akiyama}, {Aoki}, {Doi}, {Fujita}, {Furusawa},
  {Hayashino}, {Imanishi}, {Iwamuro}, {Iye}, {Kawabata}, {Kobayashi}, {Kodama},
  {Komiyama}, {Kosugi}, {Matsuda}, {Miyazaki}, {Mizumoto}, {Motohara},
  {Murayama}, {Nagao}, {Nariai}, {Ohta}, {Ohyama}, {Okamura}, {Ouchi},
  {Sasaki}, {Sekiguchi}, {Shimasaku}, {Shioya}, {Takata}, {Tamura}, {Terada},
  {Umemura}, {Usuda}, {Yagi}, {Yamada}, {Yasuda}, \& {Yoshida}}]{Kodaira2003}
{Kodaira}, K., {Taniguchi}, Y., {Kashikawa}, N., {et~al.} 2003, Publications of
  the Astronomical Society of Japan, 55, L17, \dodoi{10.1093/pasj/55.2.L17}

\bibitem[{{Konno} {et~al.}(2016){Konno}, {Ouchi}, {Nakajima}, {Duval},
  {Kusakabe}, {Ono}, \& {Shimasaku}}]{Konno2016}
{Konno}, A., {Ouchi}, M., {Nakajima}, K., {et~al.} 2016, \apj, 823, 20,
  \dodoi{10.3847/0004-637X/823/1/20}

\bibitem[{Kudritzki {et~al.}(2000)Kudritzki, Méndez, Feldmeier, Ciardullo,
  Jacoby, Freeman, Arnaboldi, Capaccioli, Gerhard, \& Ford}]{Kudritzki2000}
Kudritzki, R.-P., Méndez, R.~H., Feldmeier, J.~J., {et~al.} 2000, The
  Astrophysical Journal, 536, 19.
\newblock \url{http://stacks.iop.org/0004-637X/536/i=1/a=19}

\bibitem[{{Lai} {et~al.}(2008){Lai}, {Huang}, {Fazio}, {Gawiser}, {Ciardullo},
  {Damen}, {Franx}, {Gronwall}, {Labbe}, {Magdis}, \& {van Dokkum}}]{Lai2008}
{Lai}, K., {Huang}, J.-S., {Fazio}, G., {et~al.} 2008, \apj, 674,
  \dodoi{10.1086/524702}

\bibitem[{{Laursen}(2010)}]{Laursen2010}
{Laursen}, P. 2010, ArXiv e-prints.
\newblock \doarXiv{1012.3175}

\bibitem[{{Law} {et~al.}(2012){Law}, {Steidel}, {Shapley}, {Nagy}, {Reddy}, \&
  {Erb}}]{Law2012}
{Law}, D.~R., {Steidel}, C.~C., {Shapley}, A.~E., {et~al.} 2012, \apj, 759, 29,
  \dodoi{10.1088/0004-637X/759/1/29}

\bibitem[{{Lee} {et~al.}(2014){Lee}, {Dey}, {Hong}, {Reddy}, {Wilson},
  {Jannuzi}, {Inami}, \& {Gonzalez}}]{Lee2014}
{Lee}, K.-S., {Dey}, A., {Hong}, S., {et~al.} 2014, \apj, 796, 126,
  \dodoi{10.1088/0004-637X/796/2/126}

\bibitem[{{Lemaux} {et~al.}(2018){Lemaux}, {Le F{\`e}vre}, {Cucciati},
  {Ribeiro}, {Tasca}, {Zamorani}, {Ilbert}, {Thomas}, {Bardelli}, {Cassata},
  {Hathi}, {Pforr}, {Smol{\v{c}}i{\'c}}, {Delvecchio}, {Novak}, {Berta},
  {McCracken}, {Koekemoer}, {Amor{\'\i}n}, {Garilli}, {Maccagni}, {Schaerer},
  \& {Zucca}}]{Lemaux2018}
{Lemaux}, B.~C., {Le F{\`e}vre}, O., {Cucciati}, O., {et~al.} 2018, \aap, 615,
  A77, \dodoi{10.1051/0004-6361/201730870}

\bibitem[{{Madau} \& {Dickinson}(2014)}]{MadauDickinson2014}
{Madau}, P., \& {Dickinson}, M. 2014, \araa, 52, 415,
  \dodoi{10.1146/annurev-astro-081811-125615}

\bibitem[{{Malhotra} \& {Rhoads}(2002)}]{MalhotraRhoads2002}
{Malhotra}, S., \& {Rhoads}, J.~E. 2002, \apj, 565, L71, \dodoi{10.1086/338980}

\bibitem[{{Marinoni} {et~al.}(2002){Marinoni}, {Davis}, {Newman}, \&
  {Coil}}]{Marinoni2002}
{Marinoni}, C., {Davis}, M., {Newman}, J.~A., \& {Coil}, A.~L. 2002, \apj, 580,
  122, \dodoi{10.1086/343092}

\bibitem[{{Matsuda} {et~al.}(2005){Matsuda}, {Yamada}, {Hayashino}, {Tamura},
  {Yamauchi}, {Murayama}, {Nagao}, {Ohta}, {Okamura}, {Ouchi}, {Shimasaku},
  {Shioya}, \& {Taniguchi}}]{matsuda05}
{Matsuda}, Y., {Yamada}, T., {Hayashino}, T., {et~al.} 2005, \apjl, 634, L125,
  \dodoi{10.1086/499071}

\bibitem[{{Matsuda} {et~al.}(2012){Matsuda}, {Yamada}, {Hayashino}, {Yamauchi},
  {Nakamura}, {Morimoto}, {Ouchi}, {Ono}, {Umemura}, \& {Mori}}]{matsuda12}
---. 2012, \mnras, 425, 878, \dodoi{10.1111/j.1365-2966.2012.21143.x}

\bibitem[{{Miller} {et~al.}(2018){Miller}, {Chapman}, {Aravena}, {Ashby},
  {Hayward}, {Vieira}, {Wei{\ss}}, {Babul}, {B{\'e}thermin}, {Bradford},
  {Brodwin}, {Carlstrom}, {Chen}, {Cunningham}, {De Breuck}, {Gonzalez},
  {Greve}, {Harnett}, {Hezaveh}, {Lacaille}, {Litke}, {Ma}, {Malkan},
  {Marrone}, {Morningstar}, {Murphy}, {Narayanan}, {Pass}, {Perry}, {Phadke},
  {Rennehan}, {Rotermund}, {Simpson}, {Spilker}, {Sreevani}, {Stark},
  {Strandet}, \& {Strom}}]{Miller2018}
{Miller}, T.~B., {Chapman}, S.~C., {Aravena}, M., {et~al.} 2018, \nat, 556,
  469, \dodoi{10.1038/s41586-018-0025-2}

\bibitem[{{Nilsson} {et~al.}(2009{\natexlab{a}}){Nilsson},
  {M{\"o}ller-Nilsson}, {M{\o}ller}, {Fynbo}, \& {Shapley}}]{Nilsson2009b}
{Nilsson}, K.~K., {M{\"o}ller-Nilsson}, O., {M{\o}ller}, P., {Fynbo}, J.~P.~U.,
  \& {Shapley}, A.~E. 2009{\natexlab{a}}, \mnras, 400, 232,
  \dodoi{10.1111/j.1365-2966.2009.15439.x}

\bibitem[{{Nilsson} {et~al.}(2009{\natexlab{b}}){Nilsson}, {Tapken},
  {M{\o}ller}, {Freudling}, {Fynbo}, {Meisenheimer}, {Laursen}, \&
  {{\"O}stlin}}]{Nilsson2009a}
{Nilsson}, K.~K., {Tapken}, C., {M{\o}ller}, P., {et~al.} 2009{\natexlab{b}},
  \aap, 498, 13, \dodoi{10.1051/0004-6361/200810881}

\bibitem[{{Nilsson} {et~al.}(2007){Nilsson}, {M{\o}ller}, {M{\"o}ller},
  {Fynbo}, {Micha{\l}owski}, {Watson}, {Ledoux}, {Rosati}, {Pedersen}, \&
  {Grove}}]{Nilsson2007}
{Nilsson}, K.~K., {M{\o}ller}, P., {M{\"o}ller}, O., {et~al.} 2007, \aap, 471,
  71, \dodoi{10.1051/0004-6361:20066949}

\bibitem[{{Ning} {et~al.}(2020){Ning}, {Jiang}, {Zheng}, {Wu}, {Bian}, {Egami},
  {Fan}, {Ho}, {Shen}, {Wang}, \& {Wu}}]{Ning2020redshift57}
{Ning}, Y., {Jiang}, L., {Zheng}, Z.-Y., {et~al.} 2020, \apj, 903, 4,
  \dodoi{10.3847/1538-4357/abb705}

\bibitem[{Ono {et~al.}(2010)Ono, Ouchi, Shimasaku, Dunlop, Farrah, McLure, \&
  Okamura}]{Ono2010}
Ono, Y., Ouchi, M., Shimasaku, K., {et~al.} 2010, The Astrophysical Journal,
  724, 1524.
\newblock \url{http://stacks.iop.org/0004-637X/724/i=2/a=1524}

\bibitem[{{Ouchi} {et~al.}(2003){Ouchi}, {Shimasaku}, {Furusawa}, {Miyazaki},
  {Doi}, {Hamabe}, {Hayashino}, {Kimura}, {Kodaira}, {Komiyama}, {Matsuda},
  {Miyazaki}, {Nakata}, {Okamura}, {Sekiguchi}, {Shioya}, {Tamura},
  {Taniguchi}, {Yagi}, \& {Yasuda}}]{Ouchi2003}
{Ouchi}, M., {Shimasaku}, K., {Furusawa}, H., {et~al.} 2003, \apj, 582, 60,
  \dodoi{10.1086/344476}

\bibitem[{{Ouchi} {et~al.}(2008){Ouchi}, {Shimasaku}, {Akiyama}, {Simpson},
  {Saito}, {Ueda}, {Furusawa}, {Sekiguchi}, {Yamada}, {Kodama}, {Kashikawa},
  {Okamura}, {Iye}, {Takata}, {Yoshida}, \& {Yoshida}}]{Ouchi2008}
{Ouchi}, M., {Shimasaku}, K., {Akiyama}, M., {et~al.} 2008, \apjs, 176, 301,
  \dodoi{10.1086/527673}

\bibitem[{{Ouchi} {et~al.}(2018){Ouchi}, {Harikane}, {Shibuya}, {Shimasaku},
  {Taniguchi}, {Konno}, {Kobayashi}, {Kajisawa}, {Nagao}, {Ono}, {Inoue},
  {Umemura}, {Mori}, {Hasegawa}, {Higuchi}, {Komiyama}, {Matsuda}, {Nakajima},
  {Saito}, \& {Wang}}]{Ouchi2018silverrush}
{Ouchi}, M., {Harikane}, Y., {Shibuya}, T., {et~al.} 2018, \pasj, 70, S13,
  \dodoi{10.1093/pasj/psx074}

\bibitem[{{Oyarz{\'u}n} {et~al.}(2017){Oyarz{\'u}n}, {Blanc}, {Gonz{\'a}lez},
  {Mateo}, \& {Bailey}}]{oyarzun17}
{Oyarz{\'u}n}, G.~A., {Blanc}, G.~A., {Gonz{\'a}lez}, V., {Mateo}, M., \&
  {Bailey}, III, J.~I. 2017, \apj, 843, 133, \dodoi{10.3847/1538-4357/aa7552}

\bibitem[{{Partridge} \& {Peebles}(1967)}]{PartridgePeebles1967}
{Partridge}, R.~B., \& {Peebles}, P.~J.~E. 1967, \apj, 147, 868,
  \dodoi{10.1086/149079}

\bibitem[{{Pirzkal} {et~al.}(2007){Pirzkal}, {Malhotra}, {Rhoads}, \&
  {Xu}}]{Pirzkal2007}
{Pirzkal}, N., {Malhotra}, S., {Rhoads}, J.~E., \& {Xu}, C. 2007, \apj, 667,
  49, \dodoi{10.1086/519485}

\bibitem[{{Rauch} {et~al.}(2008){Rauch}, {Haehnelt}, {Bunker}, {Becker},
  {Marleau}, {Graham}, {Cristiani}, {Jarvis}, {Lacey}, {Morris}, {Peroux},
  {R{\"o}ttgering}, \& {Theuns}}]{Rauch2008}
{Rauch}, M., {Haehnelt}, M., {Bunker}, A., {et~al.} 2008, \apj, 681, 856,
  \dodoi{10.1086/525846}

\bibitem[{{Reddy} \& {Steidel}(2009)}]{ReddySteidel2009}
{Reddy}, N.~A., \& {Steidel}, C.~C. 2009, \apj, 692, 778,
  \dodoi{10.1088/0004-637X/692/1/778}

\bibitem[{Rhoads {et~al.}(2000)Rhoads, Malhotra, Dey, Stern, Spinrad, \&
  Jannuzi}]{Rhoads2000}
Rhoads, J.~E., Malhotra, S., Dey, A., {et~al.} 2000, The Astrophysical Journal
  Letters, 545, L85.
\newblock \url{http://stacks.iop.org/1538-4357/545/i=2/a=L85}

\bibitem[{{Ribeiro} {et~al.}(2016){Ribeiro}, {Le F{\`e}vre}, {Tasca}, {Lemaux},
  {Cassata}, {Garilli}, {Maccagni}, {Zamorani}, {Zucca}, {Amor{\'\i}n},
  {Bardelli}, {Fontana}, {Giavalisco}, {Hathi}, {Koekemoer}, {Pforr}, {Tresse},
  \& {Dunlop}}]{Ribeiro2016}
{Ribeiro}, B., {Le F{\`e}vre}, O., {Tasca}, L.~A.~M., {et~al.} 2016, \aap, 593,
  A22, \dodoi{10.1051/0004-6361/201628249}

\bibitem[{{Rotermund} {et~al.}(2021){Rotermund}, {Chapman}, {Phadke}, {Hill},
  {Pass}, {Aravena}, {Ashby}, {Babul}, {B{\'e}thermin}, {Canning}, {de Breuck},
  {Dong}, {Gonzalez}, {Hayward}, {Jarugula}, {Marrone}, {Narayanan}, {Reuter},
  {Scott}, {Spilker}, {Vieira}, {Wang}, \& {Weiss}}]{Rotermund2021}
{Rotermund}, K.~M., {Chapman}, S.~C., {Phadke}, K.~A., {et~al.} 2021, \mnras,
  502, 1797, \dodoi{10.1093/mnras/stab103}

\bibitem[{{Salpeter}(1955)}]{Salpeter1955}
{Salpeter}, E.~E. 1955, \apj, 121, 161, \dodoi{10.1086/145971}

\bibitem[{{Santos} {et~al.}(2004){Santos}, {Ellis}, {Kneib}, {Richard}, \&
  {Kuijken}}]{Santos2004}
{Santos}, M.~R., {Ellis}, R.~S., {Kneib}, J.-P., {Richard}, J., \& {Kuijken},
  K. 2004, \apj, 606, 683, \dodoi{10.1086/383080}

\bibitem[{{Schechter}(1976)}]{Schechter1976}
{Schechter}, P. 1976, \apj, 203, 297, \dodoi{10.1086/154079}

\bibitem[{{Shapley} {et~al.}(2003){Shapley}, {Steidel}, {Pettini}, \&
  {Adelberger}}]{Shapley2003}
{Shapley}, A.~E., {Steidel}, C.~C., {Pettini}, M., \& {Adelberger}, K.~L. 2003,
  \apj, 588, 65, \dodoi{10.1086/373922}

\bibitem[{{Shi} {et~al.}(2019){Shi}, {Huang}, {Lee}, {Toshikawa}, {Bowen},
  {Malavasi}, {Lemaux}, {Cucciati}, {Le Fevre}, \& {Dey}}]{Shi19b}
{Shi}, K., {Huang}, Y., {Lee}, K.-S., {et~al.} 2019, \apj, 879, 9,
  \dodoi{10.3847/1538-4357/ab2118}

\bibitem[{{Shibuya} {et~al.}(2015){Shibuya}, {Ouchi}, \&
  {Harikane}}]{Shibuya2015}
{Shibuya}, T., {Ouchi}, M., \& {Harikane}, Y. 2015, The Astrophysical Journal
  Supplement Series, 219, 15, \dodoi{10.1088/0067-0049/219/2/15}

\bibitem[{{Shimakawa} {et~al.}(2018){Shimakawa}, {Kodama}, {Hayashi},
  {Prochaska}, {Tanaka}, {Cai}, {Suzuki}, {Tadaki}, \&
  {Koyama}}]{Shimakawa2018}
{Shimakawa}, R., {Kodama}, T., {Hayashi}, M., {et~al.} 2018, \mnras, 473, 1977,
  \dodoi{10.1093/mnras/stx2494}

\bibitem[{{Shimasaku} {et~al.}(2006){Shimasaku}, {Kashikawa}, {Doi}, {Ly},
  {Malkan}, {Matsuda}, {Ouchi}, {Hayashino}, {Iye}, {Motohara}, {Murayama},
  {Nagao}, {Ohta}, {Okamura}, {Sasaki}, {Shioya}, \&
  {Taniguchi}}]{Shimasaku2006}
{Shimasaku}, K., {Kashikawa}, N., {Doi}, M., {et~al.} 2006, Publications of the
  Astronomical Society of Japan, 58, 313, \dodoi{10.1093/pasj/58.2.313}

\bibitem[{{Shioya} {et~al.}(2009){Shioya}, {Taniguchi}, {Sasaki}, {Nagao},
  {Murayama}, {Saito}, {Ideue}, {Nakajima}, {Matsuoka}, {Trump}, {Scoville},
  {Sanders}, {Mobasher}, {Aussel}, {Capak}, {Kartaltepe}, {Koekemoer},
  {Carilli}, {Ellis}, {Garilli}, {Giavalisco}, {Kitzbichler}, {Impey},
  {LeFevre}, {Schinnerer}, \& {Smolcic}}]{Shioya09}
{Shioya}, Y., {Taniguchi}, Y., {Sasaki}, S.~S., {et~al.} 2009, \apj, 696, 546,
  \dodoi{10.1088/0004-637X/696/1/546}

\bibitem[{{Sobral} {et~al.}(2018){Sobral}, {Santos}, {Matthee},
  {Paulino-Afonso}, {Ribeiro}, {Calhau}, \& {Khostovan}}]{Sobral2018}
{Sobral}, D., {Santos}, S., {Matthee}, J., {et~al.} 2018, \mnras, 476, 4725,
  \dodoi{10.1093/mnras/sty378}

\bibitem[{{Sobral} {et~al.}(2017){Sobral}, {Matthee}, {Best}, {Stroe},
  {R{\"o}ttgering}, {Oteo}, {Smail}, {Morabito}, \&
  {Paulino-Afonso}}]{Sobral2017}
{Sobral}, D., {Matthee}, J., {Best}, P., {et~al.} 2017, \mnras, 466, 1242,
  \dodoi{10.1093/mnras/stw3090}

\bibitem[{{Steidel} {et~al.}(1999){Steidel}, {Adelberger}, {Giavalisco},
  {Dickinson}, \& {Pettini}}]{Steidel1999}
{Steidel}, C.~C., {Adelberger}, K.~L., {Giavalisco}, M., {Dickinson}, M., \&
  {Pettini}, M. 1999, \apj, 519, 1, \dodoi{10.1086/307363}

\bibitem[{{van Breukelen} {et~al.}(2005){van Breukelen}, {Jarvis}, \&
  {Venemans}}]{vanBreukelen2005}
{van Breukelen}, C., {Jarvis}, M.~J., \& {Venemans}, B.~P. 2005, \mnras, 359,
  895, \dodoi{10.1111/j.1365-2966.2005.08916.x}

\bibitem[{{van de Weygaert}(1994)}]{vandeWeygaert1994}
{van de Weygaert}, R. 1994, \aap, 283, 361

\bibitem[{{Vanzella} {et~al.}(2009){Vanzella}, {Giavalisco}, {Dickinson},
  {Cristiani}, {Nonino}, {Kuntschner}, {Popesso}, {Rosati}, {Renzini}, {Stern},
  {Cesarsky}, {Ferguson}, \& {Fosbury}}]{Vanzella09}
{Vanzella}, E., {Giavalisco}, M., {Dickinson}, M., {et~al.} 2009, \apj, 695,
  1163, \dodoi{10.1088/0004-637X/695/2/1163}

\bibitem[{{Verhamme} {et~al.}(2008){Verhamme}, {Schaerer}, {Atek}, \&
  {Tapken}}]{Verhamme2008}
{Verhamme}, A., {Schaerer}, D., {Atek}, H., \& {Tapken}, C. 2008, \aap, 491,
  89, \dodoi{10.1051/0004-6361:200809648}

\bibitem[{{Verhamme} {et~al.}(2006){Verhamme}, {Schaerer}, \&
  {Maselli}}]{Verhamme2006}
{Verhamme}, A., {Schaerer}, D., \& {Maselli}, A. 2006, \aap, 460, 397,
  \dodoi{10.1051/0004-6361:20065554}

\bibitem[{{Willis} {et~al.}(2006){Willis}, {Courbin}, {Kneib}, \&
  {Minniti}}]{Willis2006}
{Willis}, J., {Courbin}, F., {Kneib}, J.-P., \& {Minniti}, D. 2006, New
  Astronomy Reviews, 50, 70, \dodoi{10.1016/j.newar.2005.11.029}

\bibitem[{{Xue} {et~al.}(2017){Xue}, {Lee}, {Dey}, {Reddy}, {Hong}, {Prescott},
  {Inami}, {Jannuzi}, \& {Gonzalez}}]{Xue2017}
{Xue}, R., {Lee}, K.-S., {Dey}, A., {et~al.} 2017, \apj, 837, 172,
  \dodoi{10.3847/1538-4357/837/2/172}

\bibitem[{{Zaninetti}(1990)}]{Zaninetti1990}
{Zaninetti}, L. 1990, \aap, 233, 293

\bibitem[{{Zheng} {et~al.}(2010){Zheng}, {Cen}, {Trac}, \& {Miralda-
  Escud{\'e}}}]{Zheng2010}
{Zheng}, Z., {Cen}, R., {Trac}, H., \& {Miralda- Escud{\'e}}, J. 2010, \apj,
  716, 574, \dodoi{10.1088/0004-637X/716/1/574}

\bibitem[{{Zheng} {et~al.}(2017){Zheng}, {Wang}, {Rhoads}, {Infante},
  {Malhotra}, {Hu}, {Walker}, {Jiang}, {Jiang}, {Hibon}, {Gonzalez}, {Kong},
  {Zheng}, {Galaz}, \& {Barrientos}}]{ZhenYaZheng2017}
{Zheng}, Z.-Y., {Wang}, J., {Rhoads}, J., {et~al.} 2017, \apjl, 842, L22,
  \dodoi{10.3847/2041-8213/aa794f}

\end{thebibliography}
\bibliographystyle{aasjournal}

\end{document}